\documentclass{pasj01}

\usepackage{url}
\usepackage{lscape}

\Received{2016/07/01}
\Accepted{2016/11/07}
\Published{}
\SetRunningHead{Q. Wada et~al.}{An X-ray Study of Dwarf Novae with Suzaku}

\begin{document}

\title{A Systematic X-ray Study of the Dwarf Novae Observed with Suzaku}
\author{
Qazuya~\textsc{Wada},\altaffilmark{1,2}
Masahiro~\textsc{Tsujimoto},\altaffilmark{1}
Ken~\textsc{Ebisawa},\altaffilmark{1,2}
Takayuki~\textsc{Hayashi}\altaffilmark{3,4}
}
\altaffiltext{1}{
Japan Aerospace Exploration Agency, Institute of Space and Astronautical Science,
3-1-1 Yoshino-dai, Chuo-ku, Sagamihara, Kanagawa 252-5210, Japan
}
\email{wada@astro.isas.jaxa.jp}
\altaffiltext{2}{
Department of Astronomy, Graduate School of Science,
The University of Tokyo, 7-3-1 Hongo, Bunkyo-ku, Tokyo 113-0033, Japan
}
\altaffiltext{3}{
Goddard Space Flight Center, National Aeronautics and Space Administration, Greenbelt, MD 20771, USA
}
\altaffiltext{4}{
Department of Physics, Faculty of Science, Nagoya University, Furo-Cho, Chikusa-ku, Nagoya 464-8602, Japan
}
\KeyWords{binaries: close --- stars: dwarf novae ---
	stars: novae, cataclysmic variables --- X-rays: stars}

\maketitle

\begin{abstract}
X-ray behavior of the dwarf novae (DNe) outside the quiescent state has not been
fully understood. We thus assembled 21 data sets of the 15 DNe observed by
the Suzaku satellite by the end of 2013, which include spectra taken during not only
the quiescence, but also the transitional, outburst, and super-outburst states.
Starting with the traditional cooling flow model to explain the X-ray emission from
the boundary layer, we made several modifications to account for the observed
spectra. As a result, we found that the best-fit spectral model depends strongly on
the state of the DNe with only a few exceptions. Spectra in the quiescent state are
explained by the cooling flow model plus a Fe fluorescent line emission attenuated by
an interstellar extinction. Spectra in the transitional state require an additional
partial covering extinction. Spectra in the outburst and super-outburst state require
additional low-temperature thin-thermal plasma component(s). Spectra in the
super-outburst state further require a high value of the minimum temperature of the
boundary layer. We present an interpretation on the required modifications to the
cooling flow model for each state.
\end{abstract}

\section{Introduction}\label{s1}
Dwarf novae (DNe) are semi-detached close binary systems consisting of a late-type
star and a white dwarf (WD) with a weak magnetic field ($\leq$10$^{6}$~G;
\cite{van-teeseling96}). The gas of the secondary star fills the Roche lobe and flows
onto the WD. The accreting gas has an angular momentum, and an accretion disk is
formed around the WD (see \cite{warner95} for a review of DNe). The accretion disk
has two stable states, in which the hydrogen is either neutral ($\sim$10$^{3}$~K) or
is fully ionized ($\sim$10$^{4}$~K). These states correspond to the quiescent and the
outburst state, respectively, in the optical light curve. The transition between the
quiescent state and the outburst state is considered to be caused by thermal
instability of the accretion disk \citep{osaki74, hoshi79, meyer81}. Some DNe also
show super-outbursts, which are brighter and longer than normal outbursts. This is
probably caused by thermal and tidal instability of the disk
\citep{whitehurst88a, osaki89, hirose90}.

These sources are also known as copious EUV and X-ray photon emitters. Whereas the
gas in the inner edge of the accretion disk follows the Kepler rotation, the rotation
velocity of the WD is much slower. Therefore, when the gas falls to the WD surface,
it is decelerated by a strong frictional force, and the kinetic energy is
dissipated into thermal energy. This region is called the boundary layer (BL), and
its temperature is heated to $\sim$10$^{8}$~K. The strong EUV and X-rays are mainly
emitted from this region.

\citet{baskill05} and \citet{pandel05} carried out a systematic study of X-ray
emission from DNe using the archival data set of the ASCA and XMM-Newton
observatories, respectively. These authors revealed that the BL consists of optically
thin multi-temperature plasma, that is explained by isobaric cooling flow model in
the quiescent state. They also reported that spectra in the outburst state deviate
from the isobaric cooling flow model. However, the number of samples in the outburst
state is small in their data sets, thus the X-ray behavior of DNe during
non-quiescent state was not fully described yet.

In this paper, we use the Suzaku X-ray archival data. A rich data set is found in the
archive, including the quiescent, transition, outburst, and super-outburst states for
the same several sources. Therefore, our catalog supplements the previous two studies
and is well suited to investigate differences in the X-ray behavior among the states.
There are two purposes of this paper; (1) to present a DNe X-ray spectral catalog
from the Suzaku archive, and (2) to describe general X-ray spectroscopic features
during each state.

\section{Observations \& Reduction}\label{s2}
\subsection{Suzaku}\label{s2-1}
We used the archival data of the Suzaku satellite \citep{mitsuda07}. Suzaku has 
two instruments: one is the X-ray Imaging Spectrometer (XIS; \cite{koyama07}) and 
the other is the Hard X-ray Detector (HXD; \cite{kokubun07, takahashi07}). We 
concentrated on the XIS data, since no significant signals were detected with HXD 
for most data sets.

The XIS consists of four X-ray CCD cameras, each of which is located in the focal
plane of the X-Ray Telescope (XRT; \cite{serlemitsos07}) modules. The XIS has a
sensitivity in an energy range of 0.2--12.0~keV band. The XIS0, 2, and 3 are
front-illuminated (FI) devices, while XIS1 is a back-illuminated (BI) device. The
former has a higher quantum efficiency in the hard band, while the latter has in the 
soft band. The combination of the XIS and the XRT offers an imaging capability to
cover a field of view of $\sim$17\farcm8$\times$17\farcm8. The entire XIS2 and a part
of XIS0 have been dysfunctional since 2006 November 9 and 2009 June 23, respectively,
due to putative micro-meteorite hits. We thus used the remaining part of XIS0, XIS1,
and XIS3 in all the data set.

\subsection{Data set}\label{s2-2}
We constructed our sample list based on \citet{ritter03}. Among those classified as
DNe, 23 sources were found to be within 9\arcmin\ of the XIS field center. We removed
sources (i) located in a globular cluster or unresolved from other sources, and
(ii) with no significant detection of the XIS by quick-look analysis (CP Dra). For
each observation, we constructed an optical light curves using the American
Association of Variable Star Observers (AAVSO)
\footnote{See \url{https://www.aavso.org/} for details.} International Database
(figure \ref{f1}). As a result, we obtained a list of 15 DNe for various types
(SU~UMa, U~Gem, Z~Cam types) with 21 observations for various states (quiescent,
outburst, super-outburst, and transitional states; table~\ref{t1}). Here, we define
the ``transitional state'' as the period between the onset and the peak, and the
``outburst state'' from the peak to the quiescence level in an outburst by the
optical light curve. Four DNe have multiple observations in different states, which
is a uniqueness of our data set.

\begin{figure*}[ht]
  \begin{minipage}{0.5\hsize}
    \begin{center}
      \includegraphics[width=80mm]{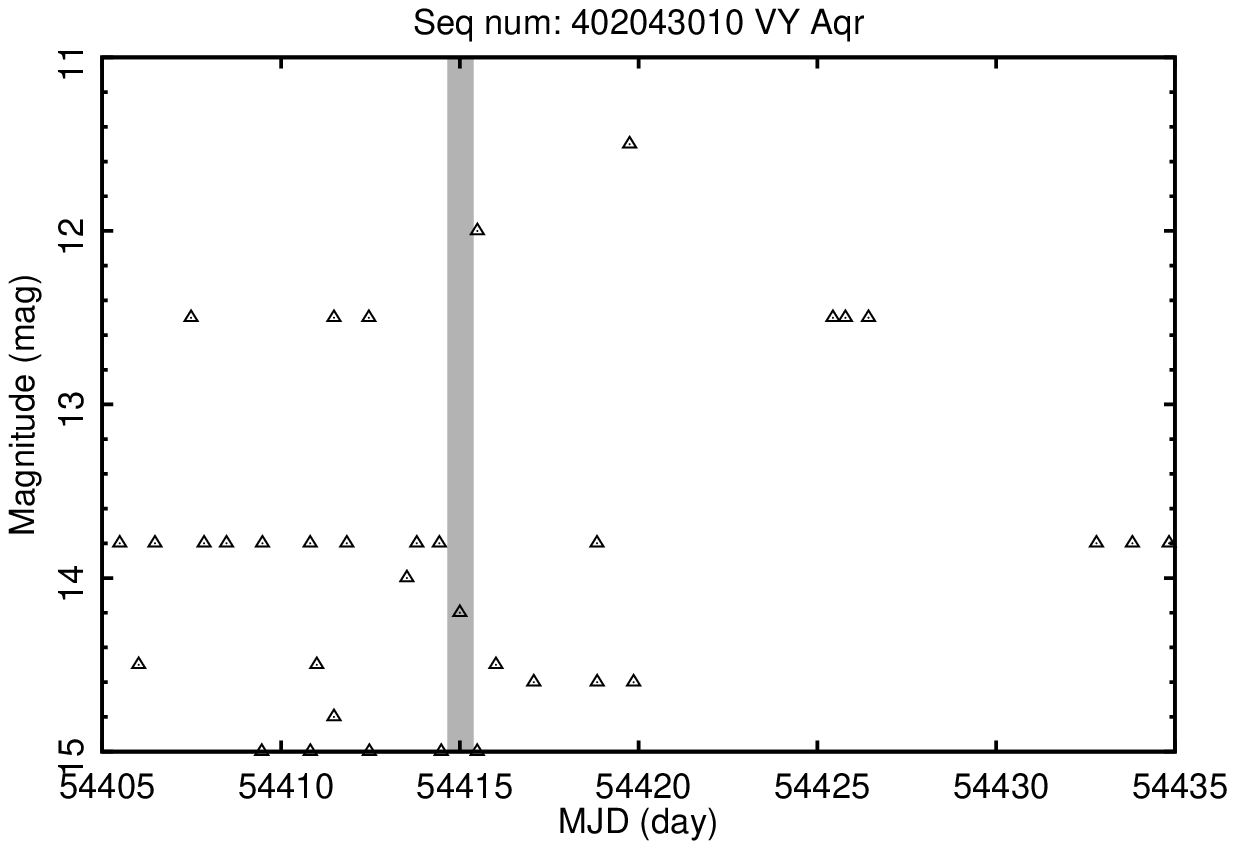}
      \includegraphics[width=80mm]{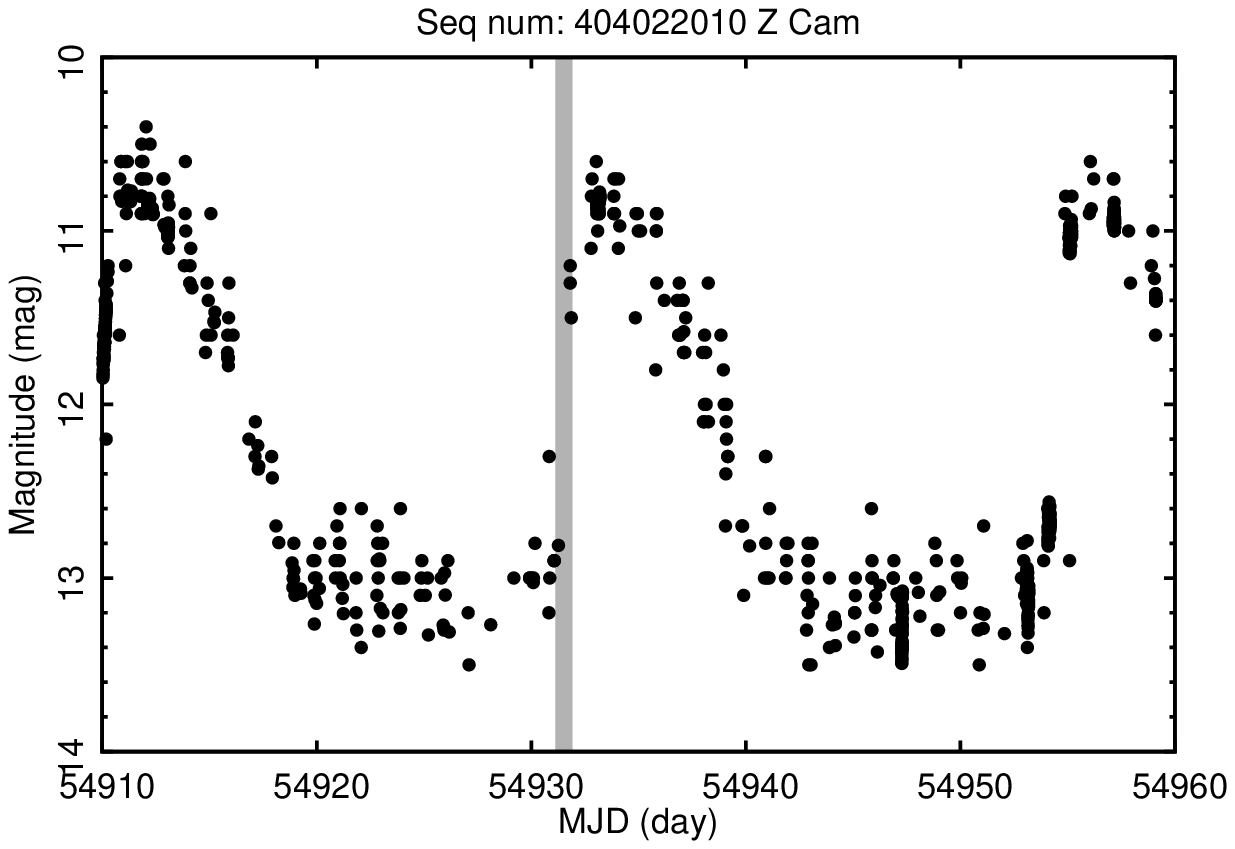}
      \includegraphics[width=80mm]{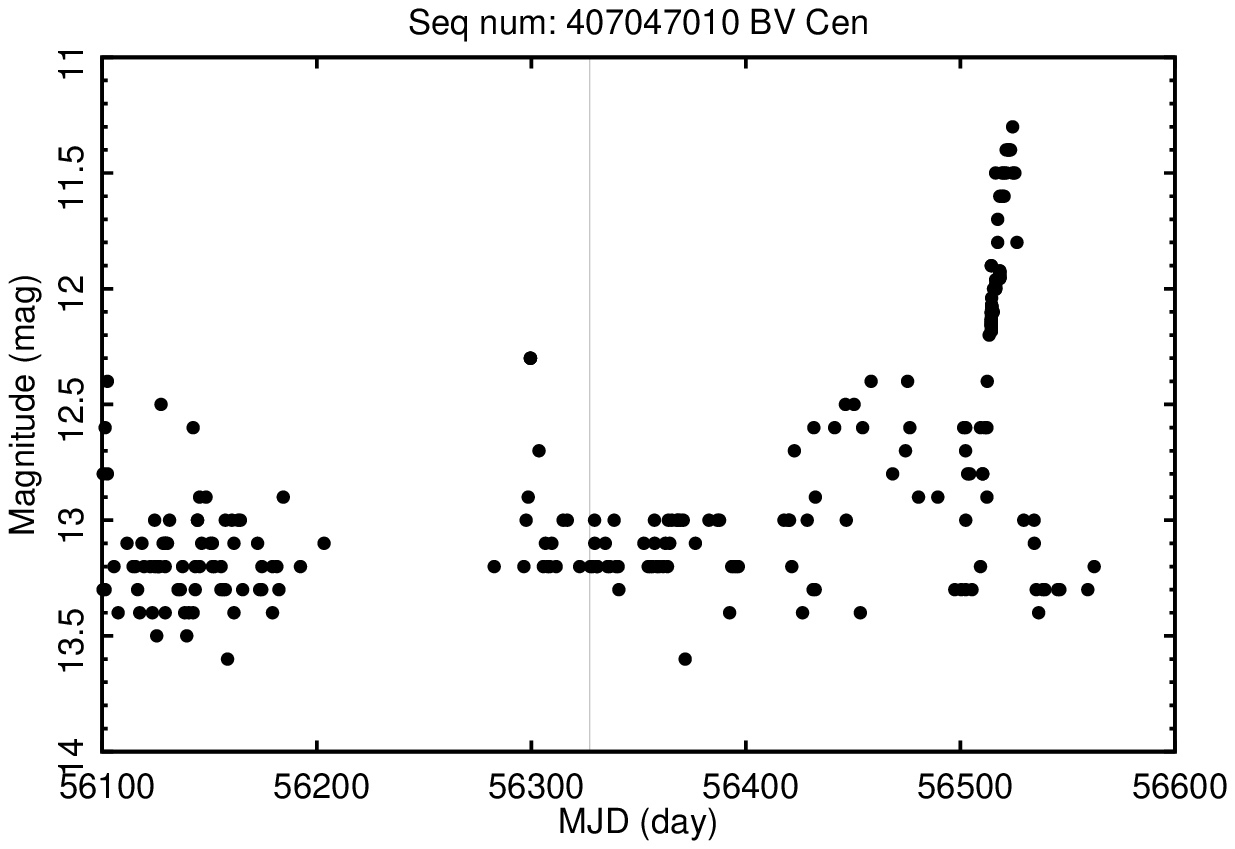}
      \includegraphics[width=80mm]{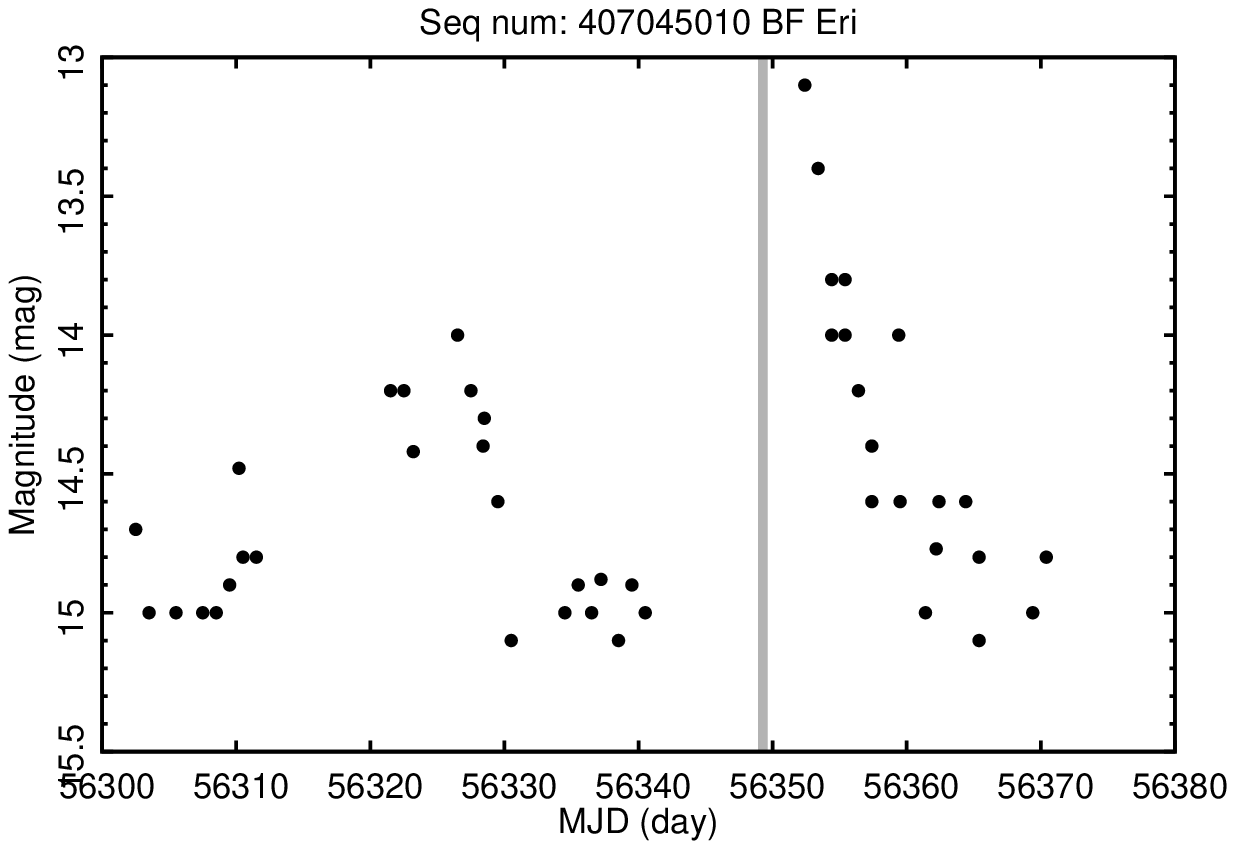}
    \end{center}
  \end{minipage}
  \begin{minipage}{0.5\hsize}
    \begin{center}
      \includegraphics[width=80mm]{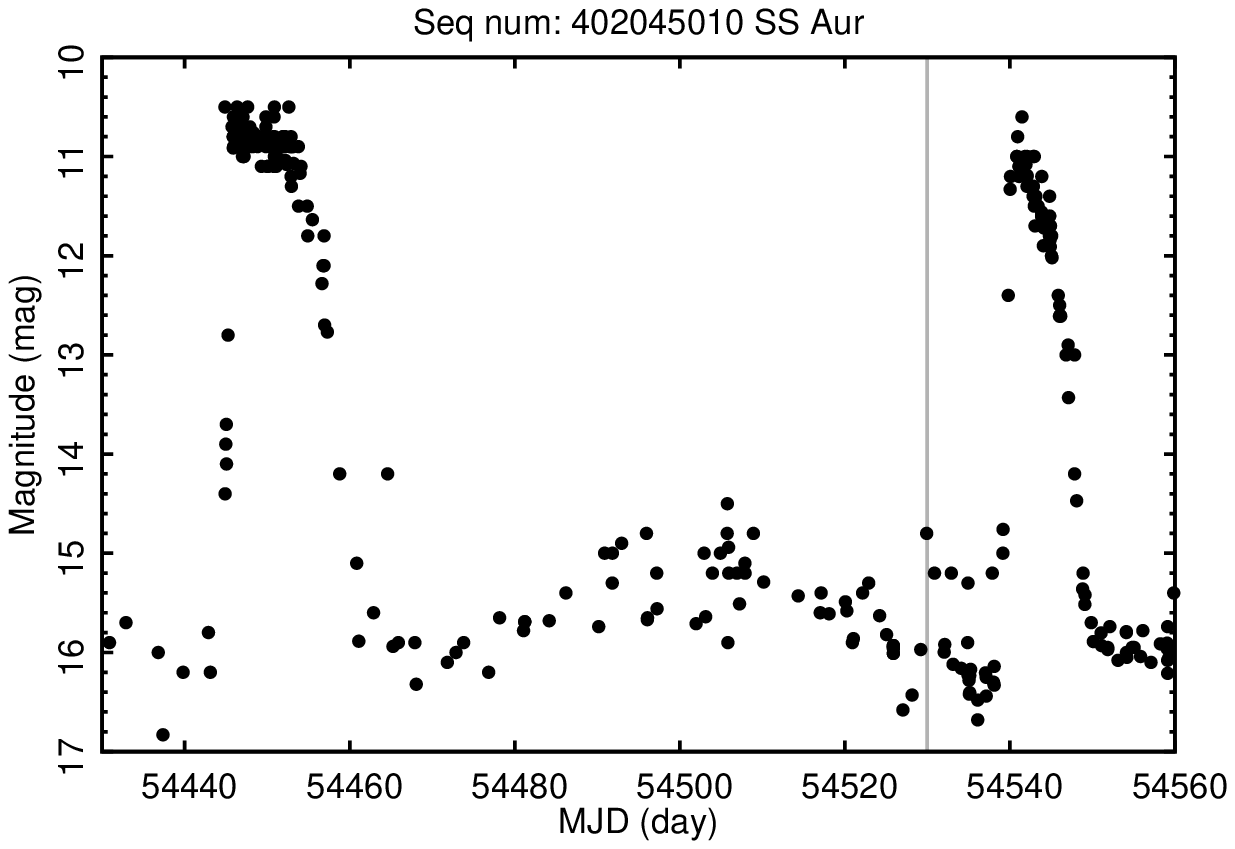}
      \includegraphics[width=80mm]{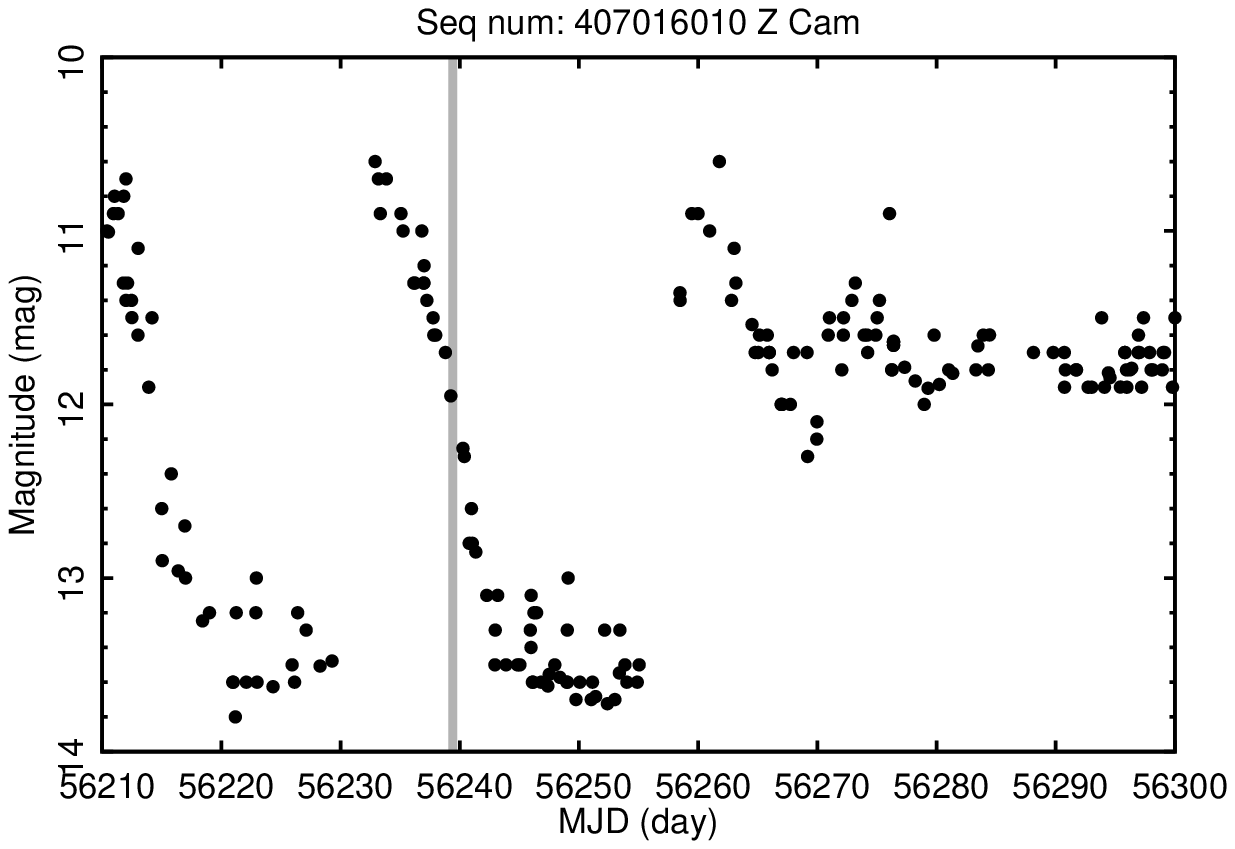}
      \includegraphics[width=80mm]{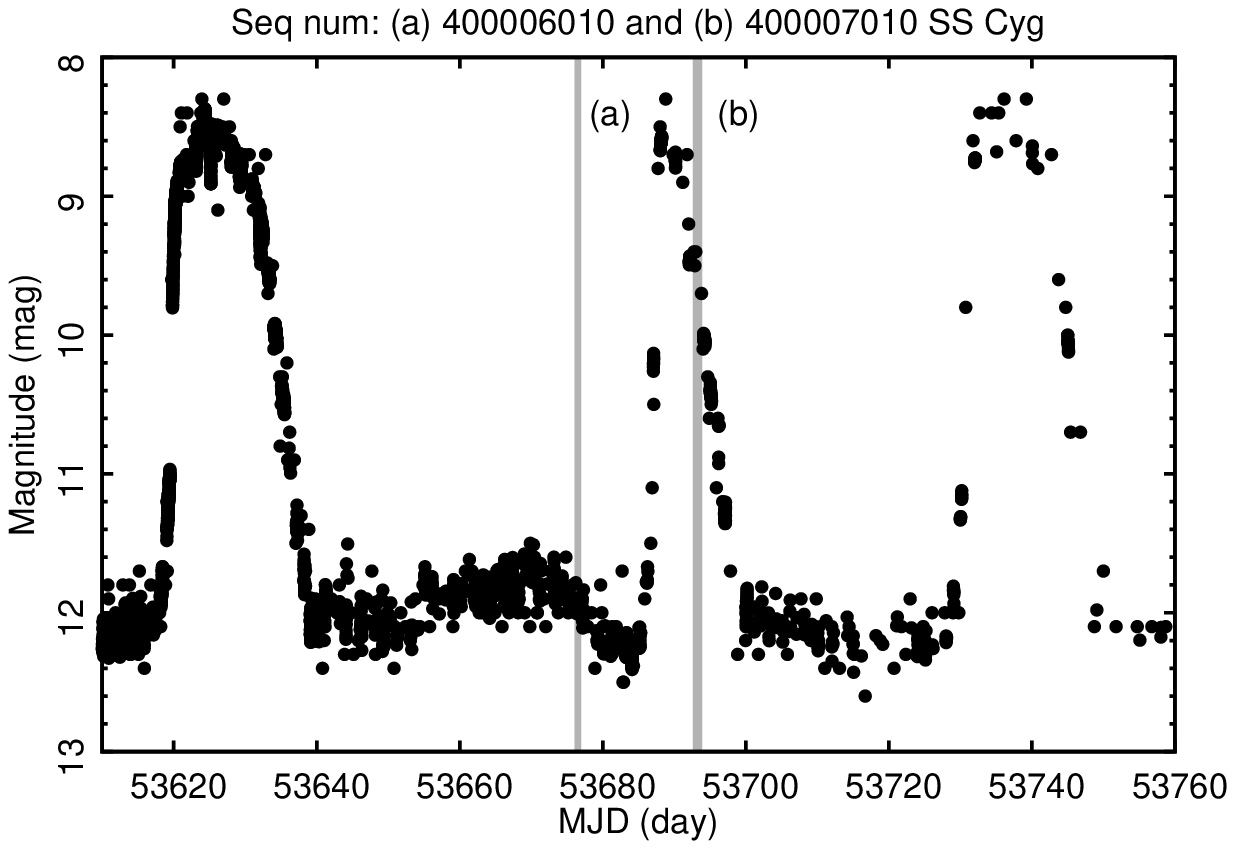}
      \includegraphics[width=80mm]{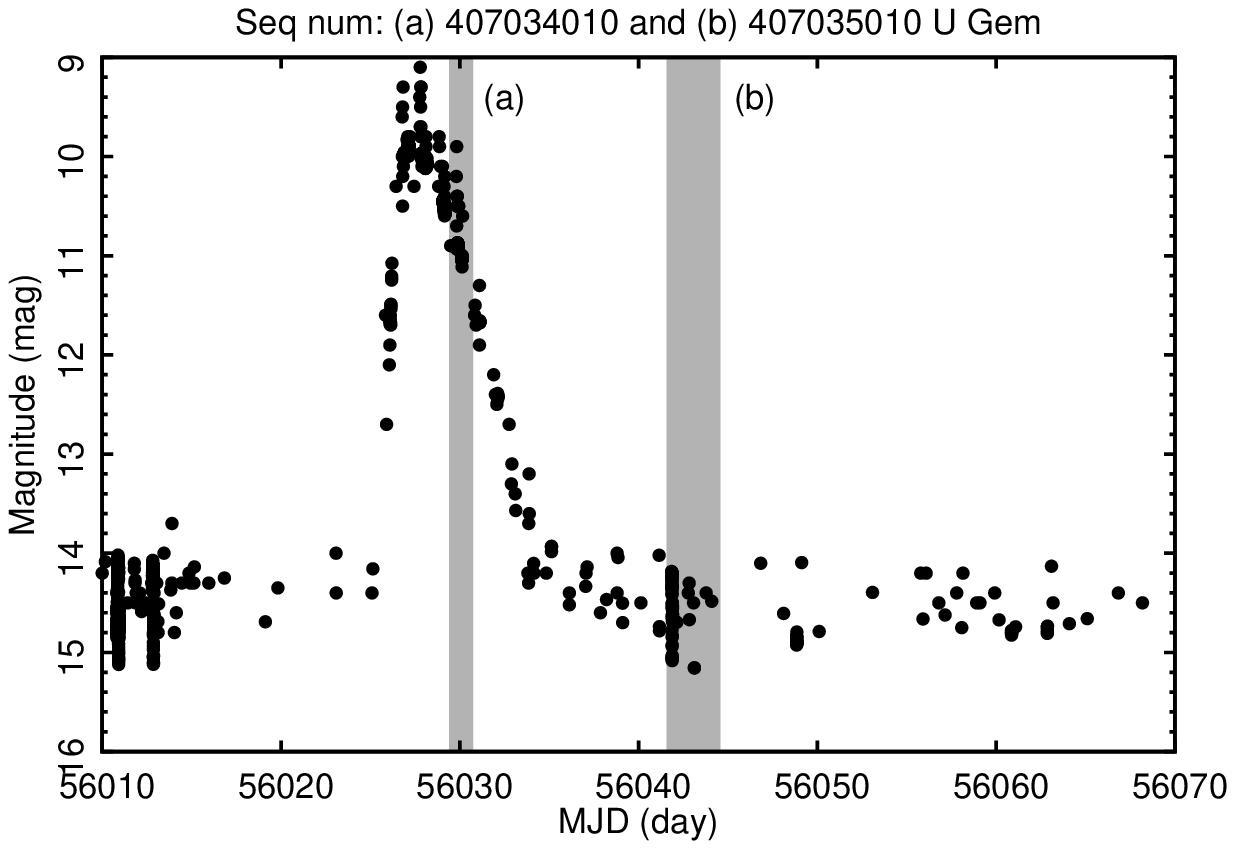}
    \end{center}
  \end{minipage}
  \vspace{1mm}
 \caption{Optical light curves from AAVSO. Filled circles are observed magnitude, 
   while open triangles are the upper limits. The hatched stripes indicate the
   duration of the Suzaku observations.}
 \label{f1}
\end{figure*}

\addtocounter{figure}{-1}
\begin{figure*}[ht]
  \begin{minipage}{0.5\hsize}
    \begin{center}
      \includegraphics[width=80mm]{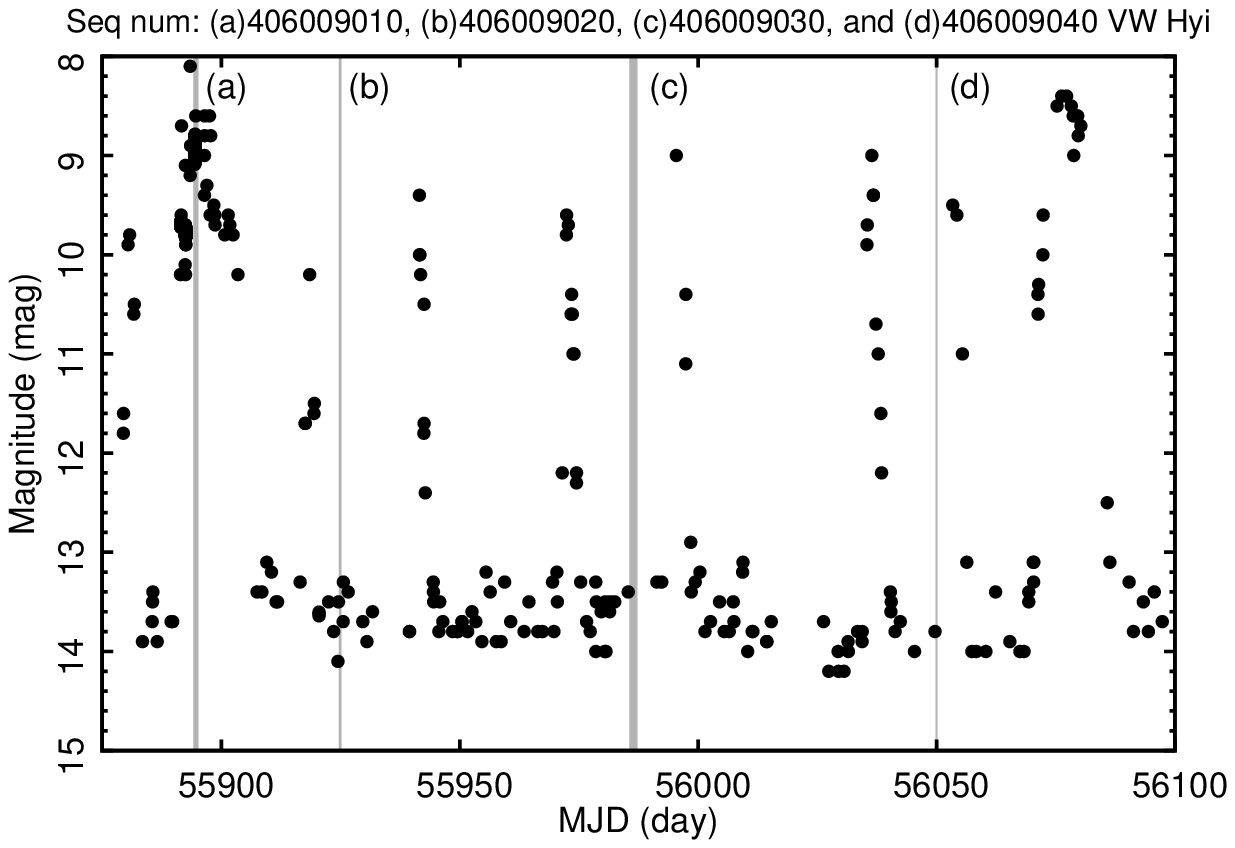}
      \includegraphics[width=80mm]{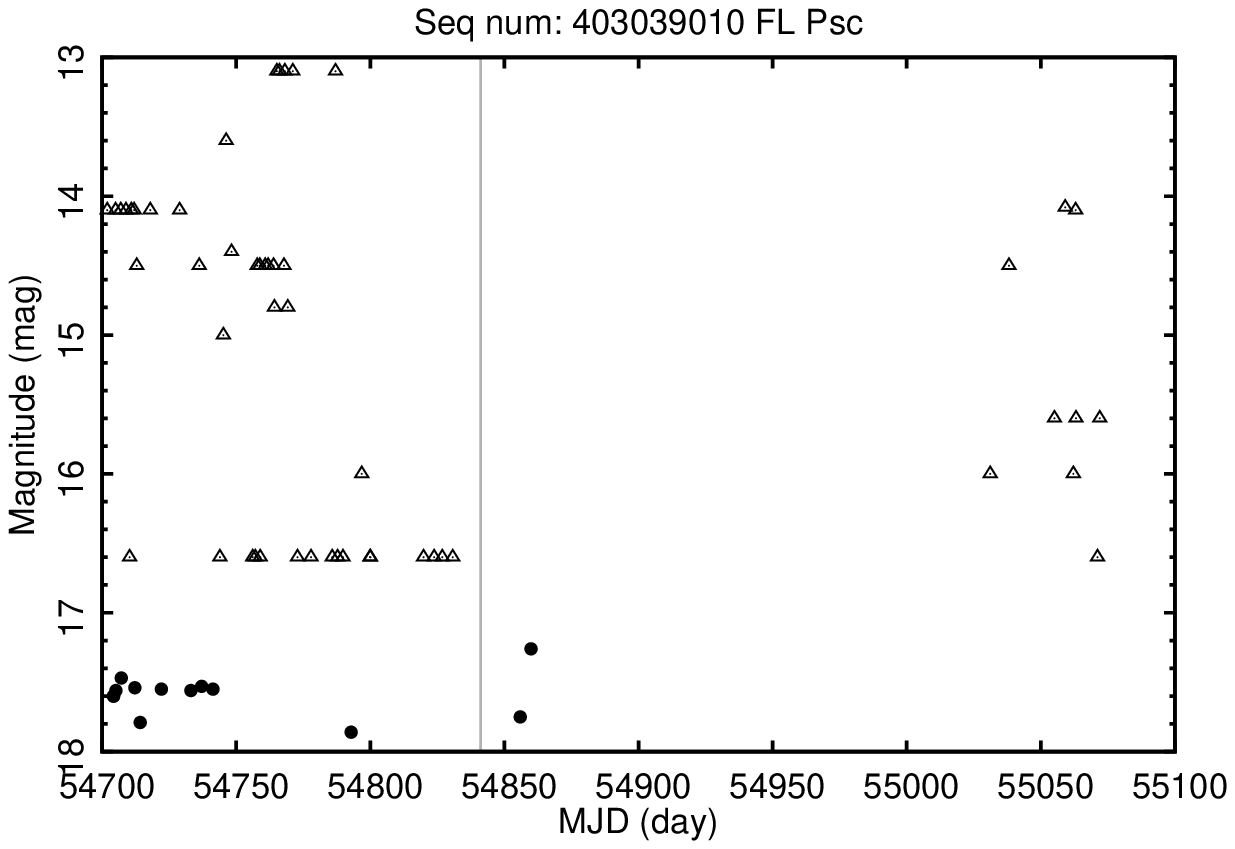}
      \includegraphics[width=80mm]{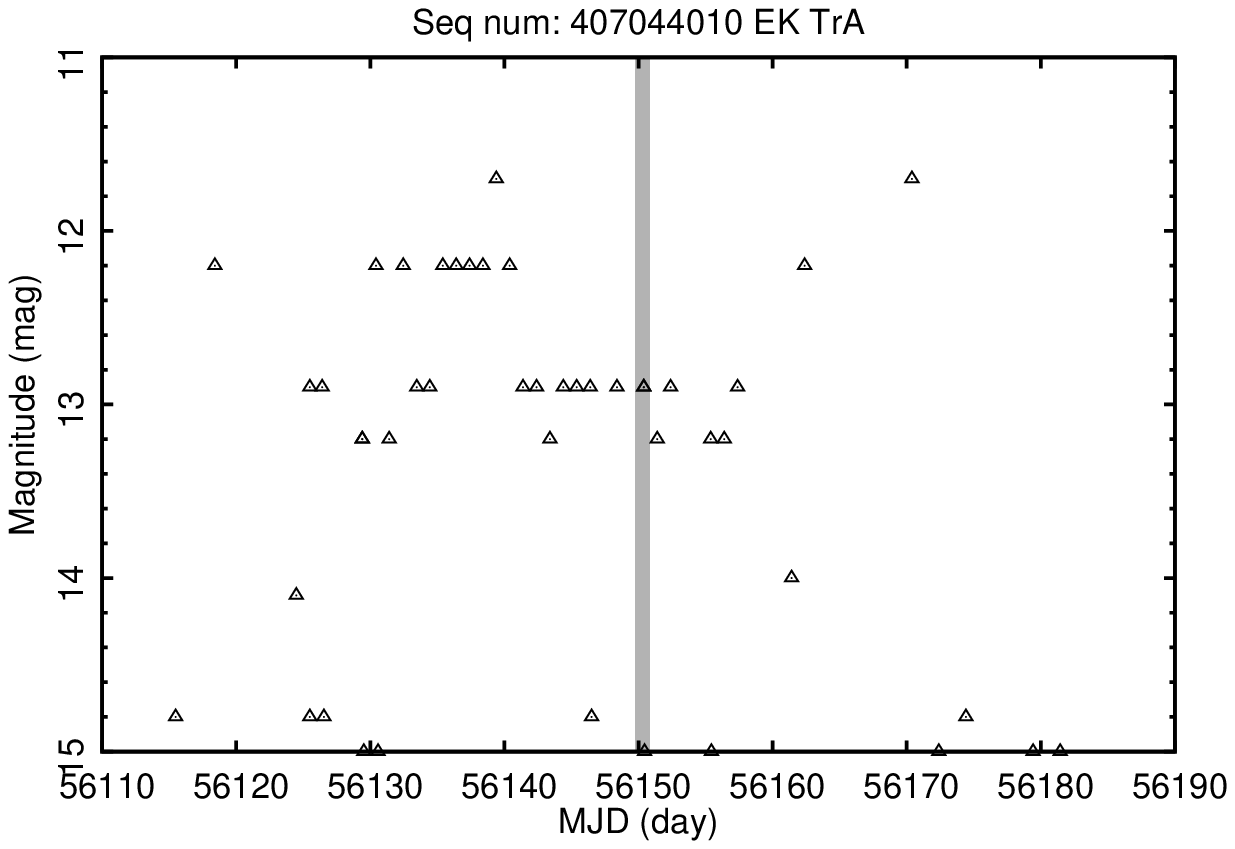}
      \includegraphics[width=80mm]{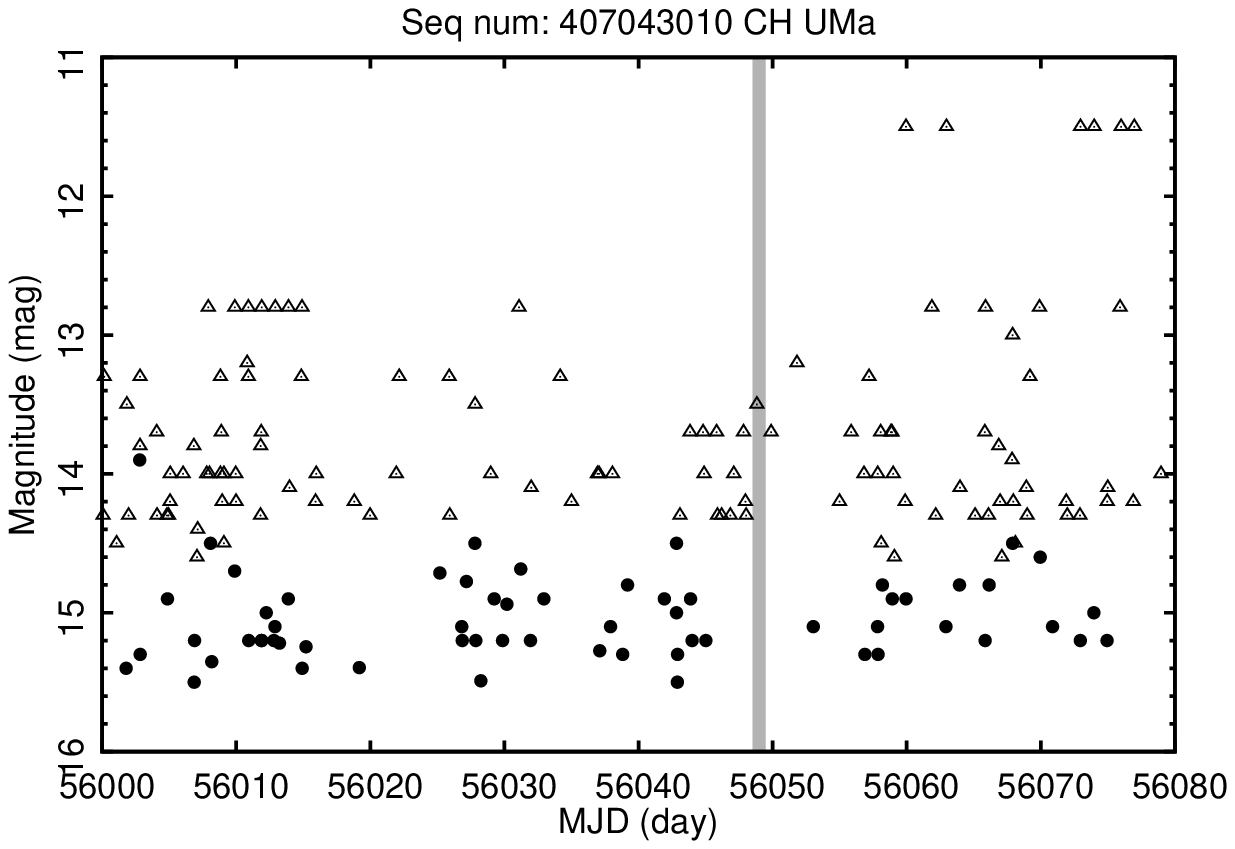}
    \end{center}
  \end{minipage}
  \begin{minipage}{0.5\hsize}
    \begin{center}
      \includegraphics[width=80mm]{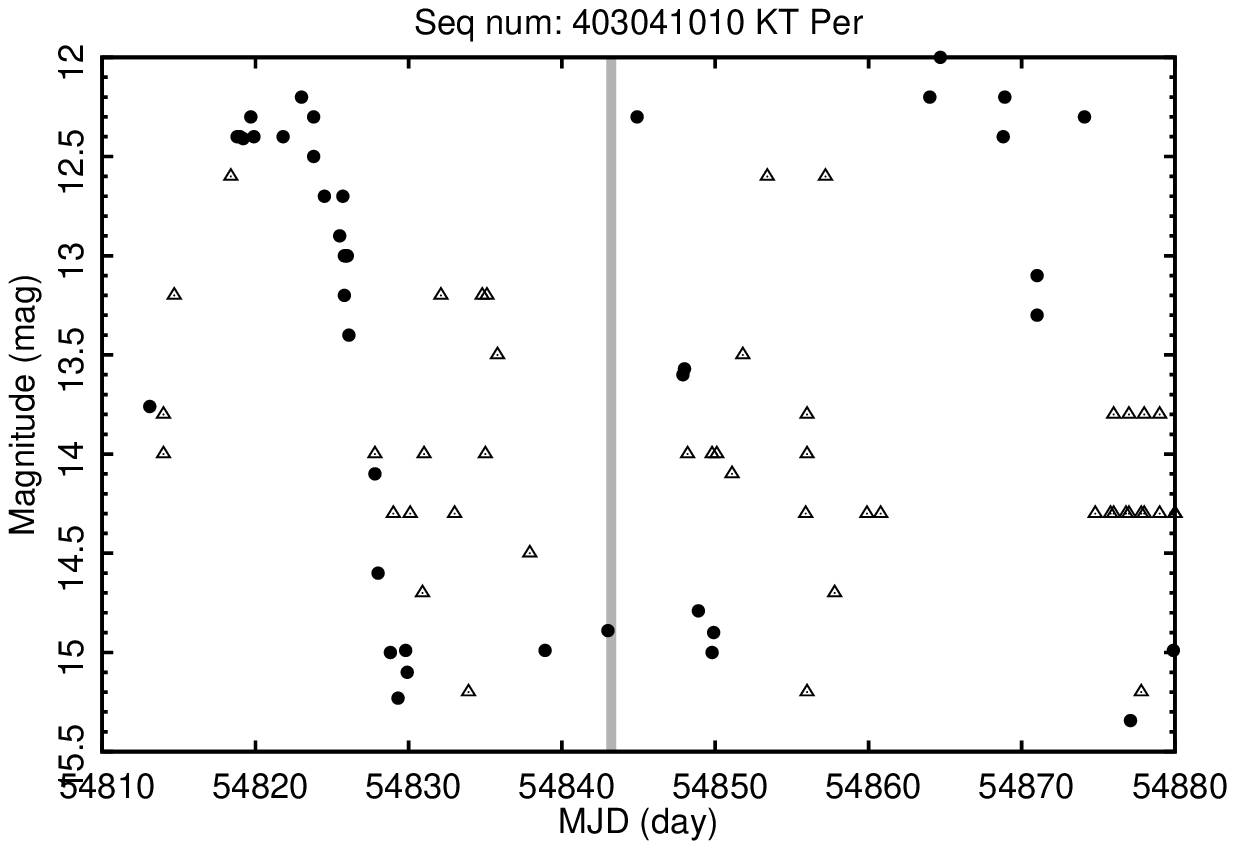}
      \includegraphics[width=80mm]{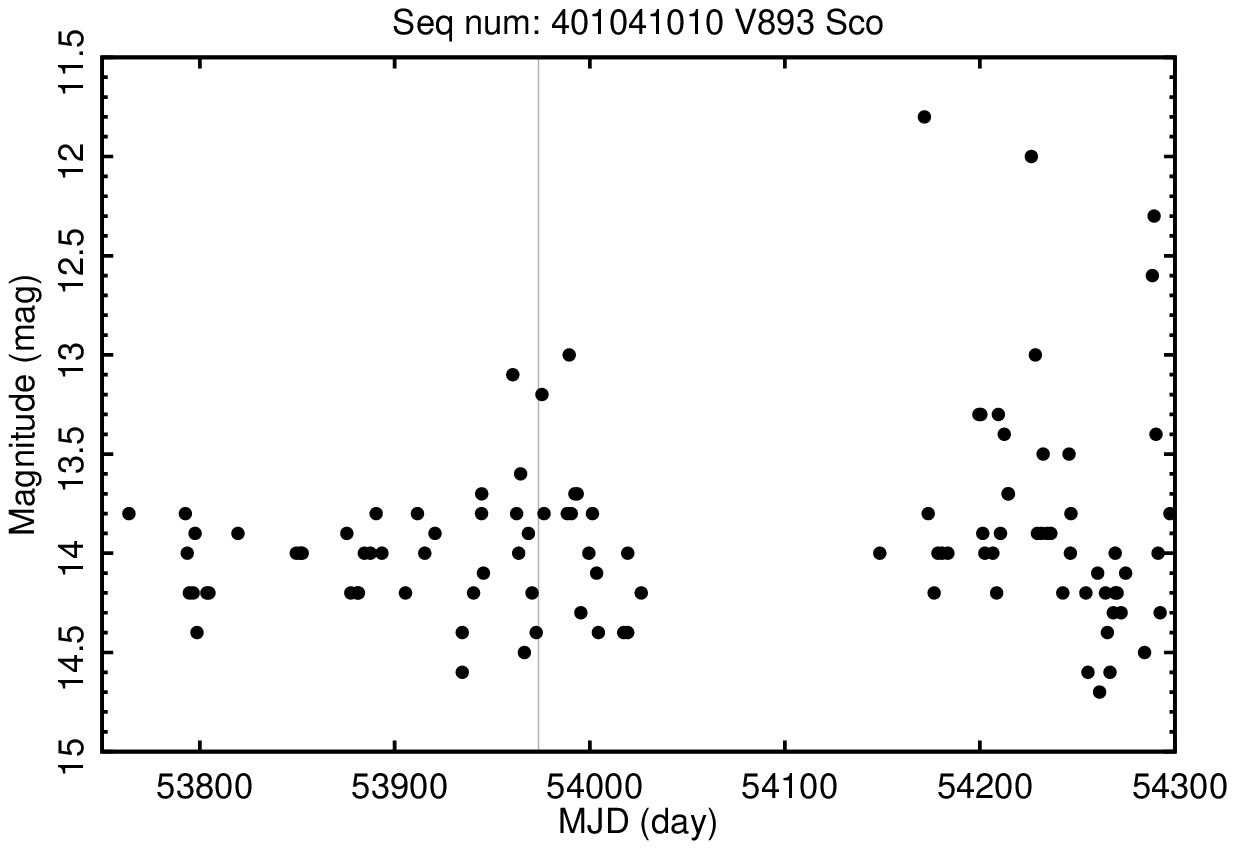}
      \includegraphics[width=80mm]{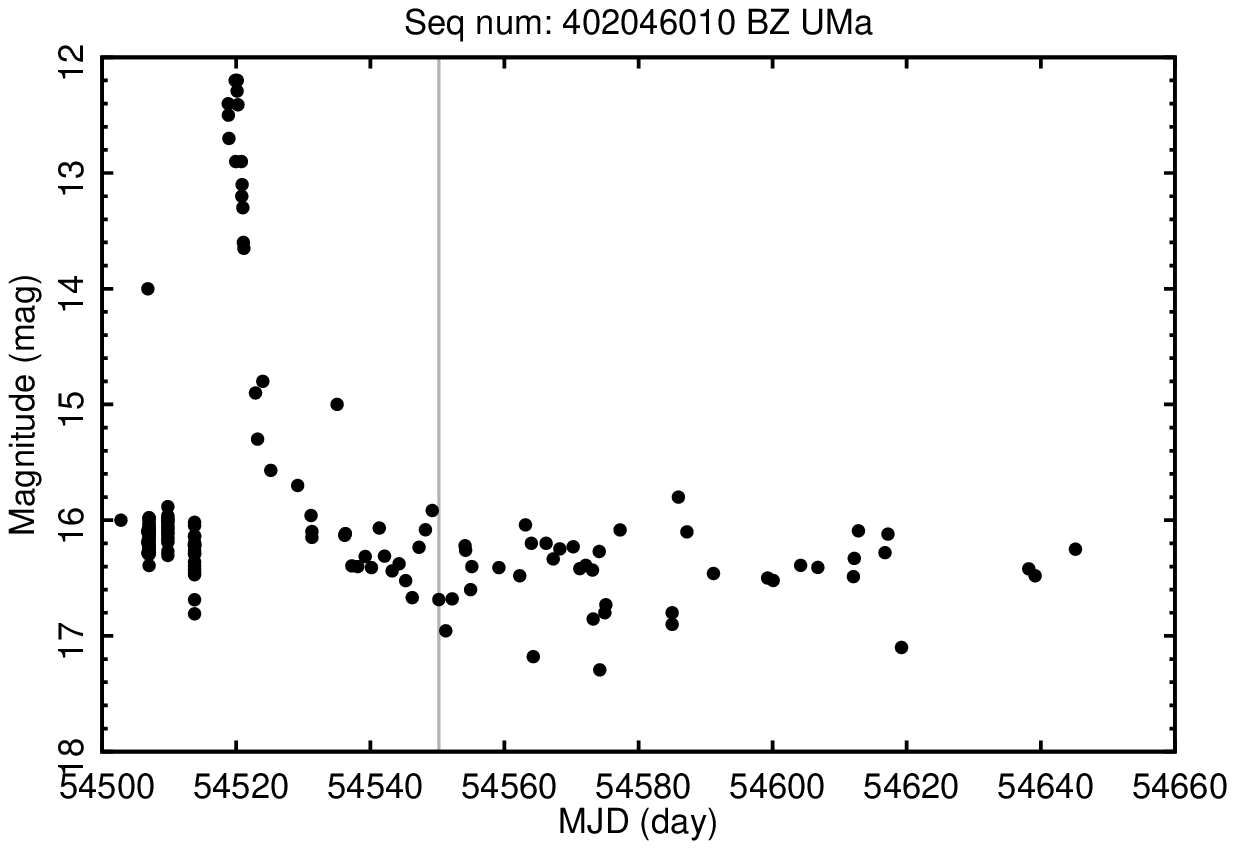}
      \includegraphics[width=80mm]{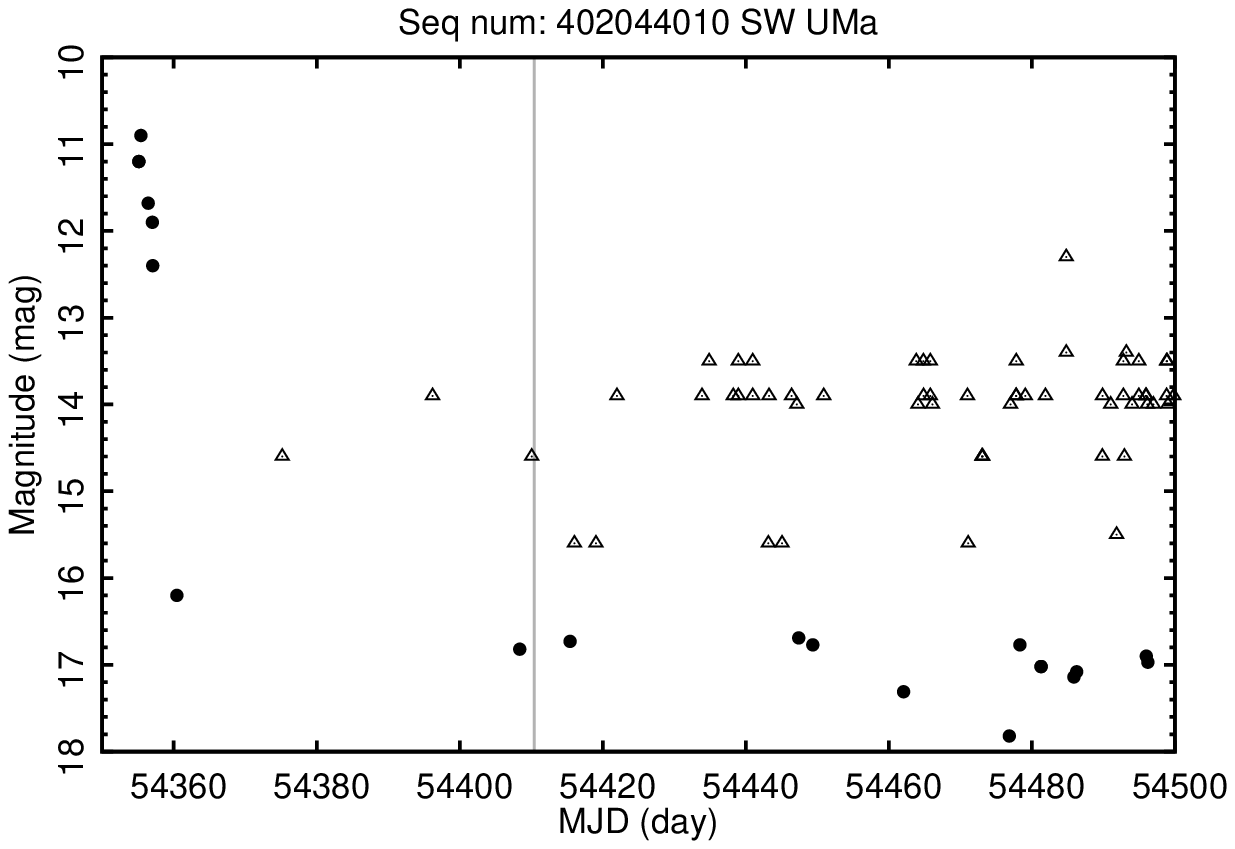}
    \end{center}
  \end{minipage}
  \vspace{1mm}
  \caption{{\it Continued.}}
\end{figure*}
\addtocounter{figure}{-1}
\begin{figure*}[ht]
  \begin{minipage}{0.5\hsize}
    \begin{center}
      \includegraphics[width=80mm]{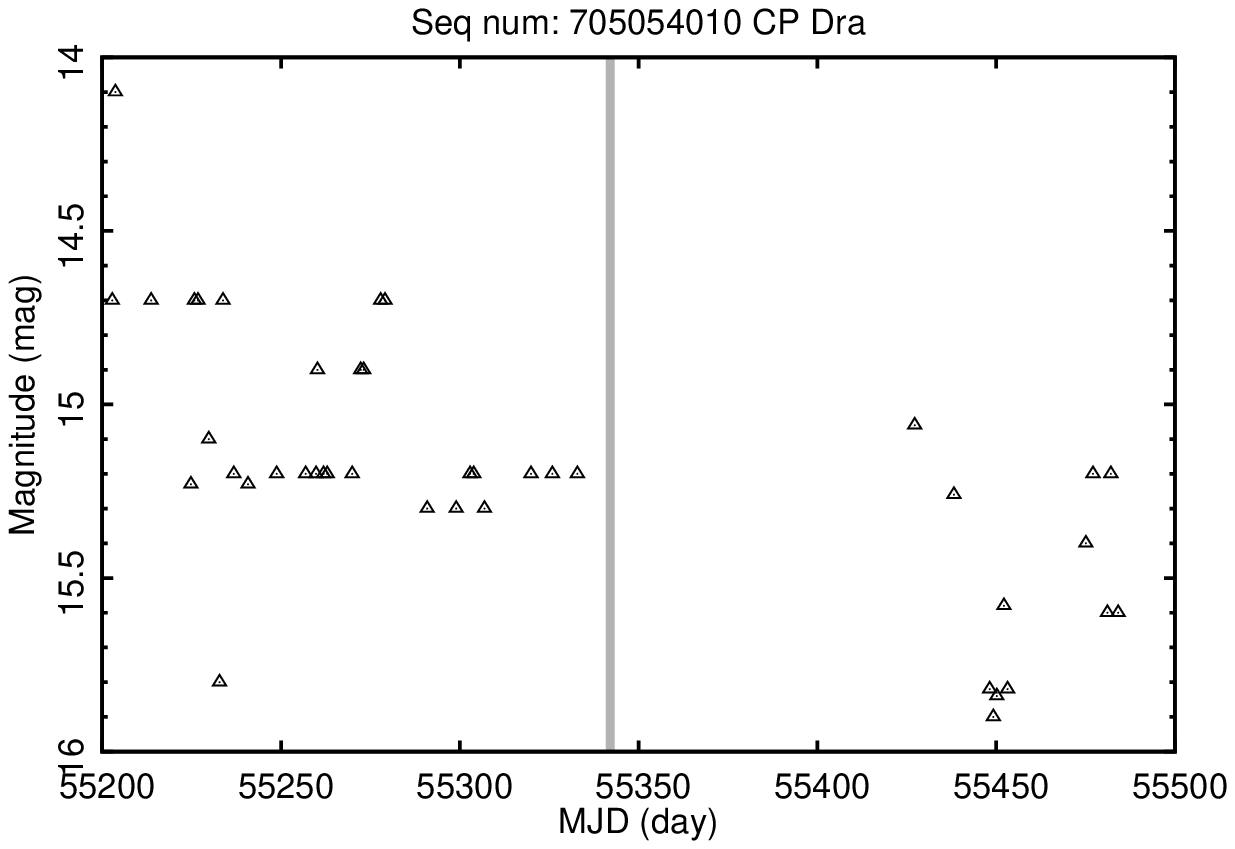}
    \end{center}
  \end{minipage}
  \vspace{1mm}
  \caption{{\it Continued.}}
\end{figure*}

\begin{table*}[ht]
  \begin{center}
    \caption{Log of Suzaku observations of dwarf novae.}
    \label{t1}
    \begin{tabular}{lccccrrccl}
      \hline
      \hline 
      Target & Type\footnotemark[$*$] &  State\footnotemark[$\dagger$] & Obs date & Seq num & \multicolumn{1}{c}{$t_{\mathrm{exp}}$}\footnotemark[$\ddagger$] & \multicolumn{1}{c}{$P_{\mathrm{orb}}$\footnotemark[$\S$]} & Inclination & Distance & References \\
      & & & & & \multicolumn{1}{c}{(ks)} &  \multicolumn{1}{c}{(hr)} & & (pc) & \\
      \hline
      VY Aqr   & SU & Q & 2007-11-10 & 402043010 & 25.4 & 1.514 & 63\degree$\pm$13\degree & 97$^{+15}_{-12}$ & 1, 2 \\
      SS Aur   & UG & Q & 2008-05-04 & 402045010 & 19.5 & 4.387 & 40\degree$\pm$7\degree & 200$\pm$26 & 3, 4, 5 \\
      Z Cam    & ZC & T & 2009-04-10 & 404022010 & 37.7 & 6.956 & 68\degree & 163$^{+68}_{-38}$ & 2, 6, 7 \\
               &    & O & 2012-11-08  & 407016010 & 35.9 \\
      BV Cen   & UG & Q & 2013-02-06 & 407047010 & 33.4 & 14.643 & 53\degree$\pm$4\degree & 400 & 8, 9, 10 \\
      SS Cyg   & UG & Q & 2005-11-02  & 400006010 & 39.5 & 6.603 & 45\degree--56\degree & 114$\pm$2 & 11, 12, 13 \\
               &    & O & 2005-11-18 & 400007010 & 56.1 \\
      BF Eri   & UG & Q & 2013-02-27 & 407045010 & 32.8 & 6.501 & 38\degree--40.5\degree & 700$\pm$200 & 14 \\
      U Gem    & UG & O & 2012-04-12 & 407035010 & 50.3 & 4.246 & 69.7\degree$\pm$0.7\degree & 96.4$\pm$4.6 & 5, 15, 16 \\
               &    & Q & 2012-04-24 & 407034010 & 119.1 \\
      VW Hyi   & SU & S & 2011-11-29 & 406009010 & 70.1 & 1.783 & 60\degree$\pm$10\degree & 82$\pm$5 & 17, 18, 19 \\
               &    & Q & 2011-12-29 & 406009020 & 16.2 \\
               &    & Q & 2012-02-29 & 406009030 & 20.1 \\
               &    & Q & 2012-05-02  & 406009040 & 16.8 \\
      KT Per   & ZC & Q or T & 2009-01-12 & 403041010 & 29.2 & 3.904 & 60\degree & 180$^{+36}_{-28}$ & 10, 20, 21 \\
      FL Psc   & SU & Q & 2009-01-10 & 403039010 & 33.3 & 1.357 & -- & 130$\pm$30 & 22, 23 \\
      V893 Sco & SU & Q & 2006-08-26 & 401041010 & 18.5 & 1.823 & 72.5\degree & 155$^{+58}_{-34}$ & 2, 24, 25 \\
      EK TrA   & SU & Q & 2012-08-10 & 407044010 & 77.9 & 1.509 & 58\degree$\pm$7\degree & 180 & 26, 27 \\
      BZ UMa   & SU & Q & 2008-03-24 & 402046010 & 29.8 & 1.632 & 57\degree & 228$^{+56}_{-38}$ & 10, 29, 30 \\
      CH UMa   & UG & Q & 2012-05-01 & 407043010 & 45.2 & 8.236 & 21\degree$\pm$4\degree & 480 & 10, 29, 30 \\
      SW UMa   & SU & Q & 2007-11-06 & 402044010 & 16.9 & 1.364 & 45\degree$\pm$18\degree & 159$\pm$22 & 4, 21, 31 \\
      CP Dra   & SU & Q & 2010-05-24 & 705054010 & 104.8 & 1.939 & --- & --- & 32 \\
      \hline
      \multicolumn{10}{l}{\parbox{160mm}{
        \footnotesize
        \par \noindent
        \footnotemark[$*$] The DNe sub-classification by \citet{ritter11}: 
        UG$=$U~Gem, ZC$=$Z~Cam, and SU$=$SU~UMa type.\\
        \footnotemark[$\dagger$] States based on the optical light curve:
        Q$=$quiescent, T$=$transitional, O$=$outburst, and S$=$super-outburst. \\
        \footnotemark[$\ddagger$] Exposure time. \\
        \footnotemark[$\S$] Orbital period.\\
        References--(1)~\cite{thorstensen97b}; (2)~\cite{thorstensen03};
        (3)~\cite{shafter86}; (4)~\cite{shafter83}; (5)~\cite{harrison99};
        (6)~\cite{hartley05}; (7)~\cite{thorstensen95}; (8)~\cite{hollander93}; 
        (9)~\cite{watson07b}; (10)~\cite{patterson11}; (11)~\cite{hessman84}; 
        (12)~\cite{bitner07}; (13)~\cite{miller-jones13}; (14)~\cite{neustroev08}
        (15)~\cite{marsh90}; (16)~\cite{zhang87}; (17)~\cite{van-amerongen87};
        (18)~\cite{schoembs81}; (19)~\cite{barrett96}; (20)~\cite{echevarria99};
        (21)~\cite{thorstensen08}; (22)~\cite{kato09}; (23)~\cite{reis13};
        (24)~\cite{mukai09}; (25)~\cite{mason01}; (26)~\cite{mennickent98};
        (27)~\cite{godon08}; (28)~\cite{jurcevic94}; (29)~\cite{thorstensen04};
        (30)~\cite{friend90}; (31)~\cite{howell88}; (32)~\cite{shears11}. \\
      }}
  \end{tabular}
 \end{center}
\end{table*}

\subsection{Data reduction}\label{s2-3}
For all observations, the XIS was operated with the normal
clocking mode with a read-out time of 8~s. We discarded events during the South
Atlantic Anomaly (SAA) passages and the elevation angles from the day Earth by
$\le$~20\degree\ and the night Earth by $\le$~5\degree. We extracted source events
within a 3\arcmin\ radius circle, and the background events from a 4--6\arcmin\
annulus for all the targets. For FL~Psc, we further removed events within a 2\arcmin\
radius circle of a near-by point-like source.

Throughout this paper, we used HEADAS software package\footnote {See
\url{http://heasarc.gsfc.nasa.gov/docs/software/heasoft/} for details.}
version 6.15 for data reduction. The response files of XIS and XRT were generated
using the {\tt xisrmfgen} and the {\tt xissimarfgen} \citep{ishisaki07} tools,
respectively.

\section{Analysis}\label{s3}
The average XIS1 spectra in the 0.25--8.0~keV band are shown for 21 observations in
figure~\ref{f2}. Almost all the spectra exhibit a complex of Fe lines in the 6--7~keV
band, which is composed of the fluorescent line from neutral or low ionized Fe,
He $\alpha$ line from Fe\emissiontype{XXV}, and Ly $\alpha$ line from
Fe\emissiontype{XXVI}. We use the 0.25--8.0 (BI) and 0.4--10.0 keV (FI) band spectra
for the fitting. We inspected all the light curves and found no clear phase change,
thus we only deal with time-averaged spectra unless otherwise noted.

\begin{figure*}[ht]
  \begin{center}
    \includegraphics[width=160mm]{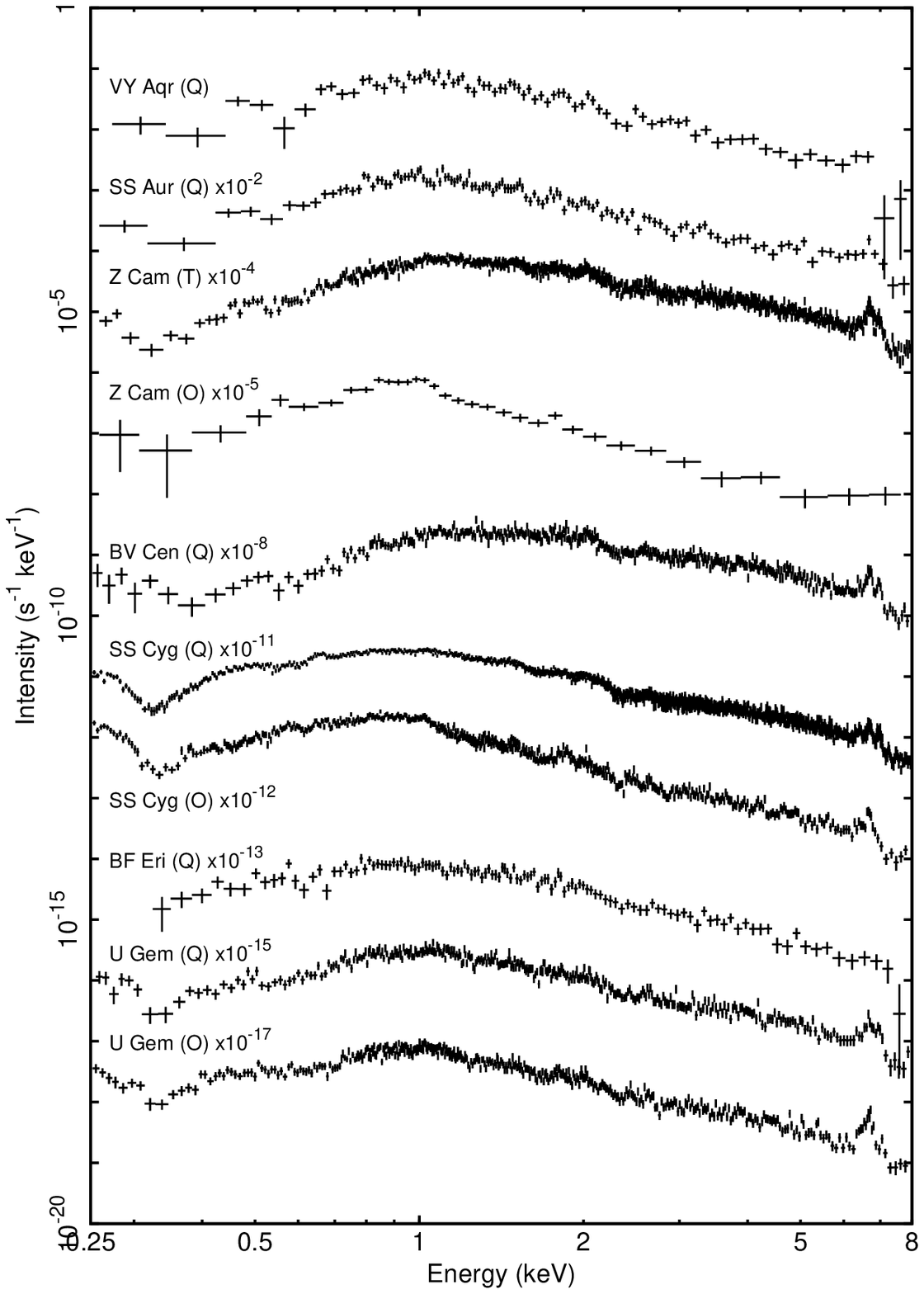}
  \end{center}
  \caption{Average XIS1 spectra in 0.25--8.0~keV band for all data sets. The spectra
    are shifted in the vertical direction for clarity. State based on the optical
    light curve: Q=quiescent, T=transitional, O=outburst, and S=super-outburst. For
    VW~Hyi in the quiescent state, Q$_{1}$, Q$_{2}$, and Q$_{3}$ correspond to the
    sequence number of 406009020, 406009030, and 406009040, respectively.
  }\label{f2}
\end{figure*}
\addtocounter{figure}{-1}
\begin{figure*}
  \begin{center}
    \includegraphics[width=160mm]{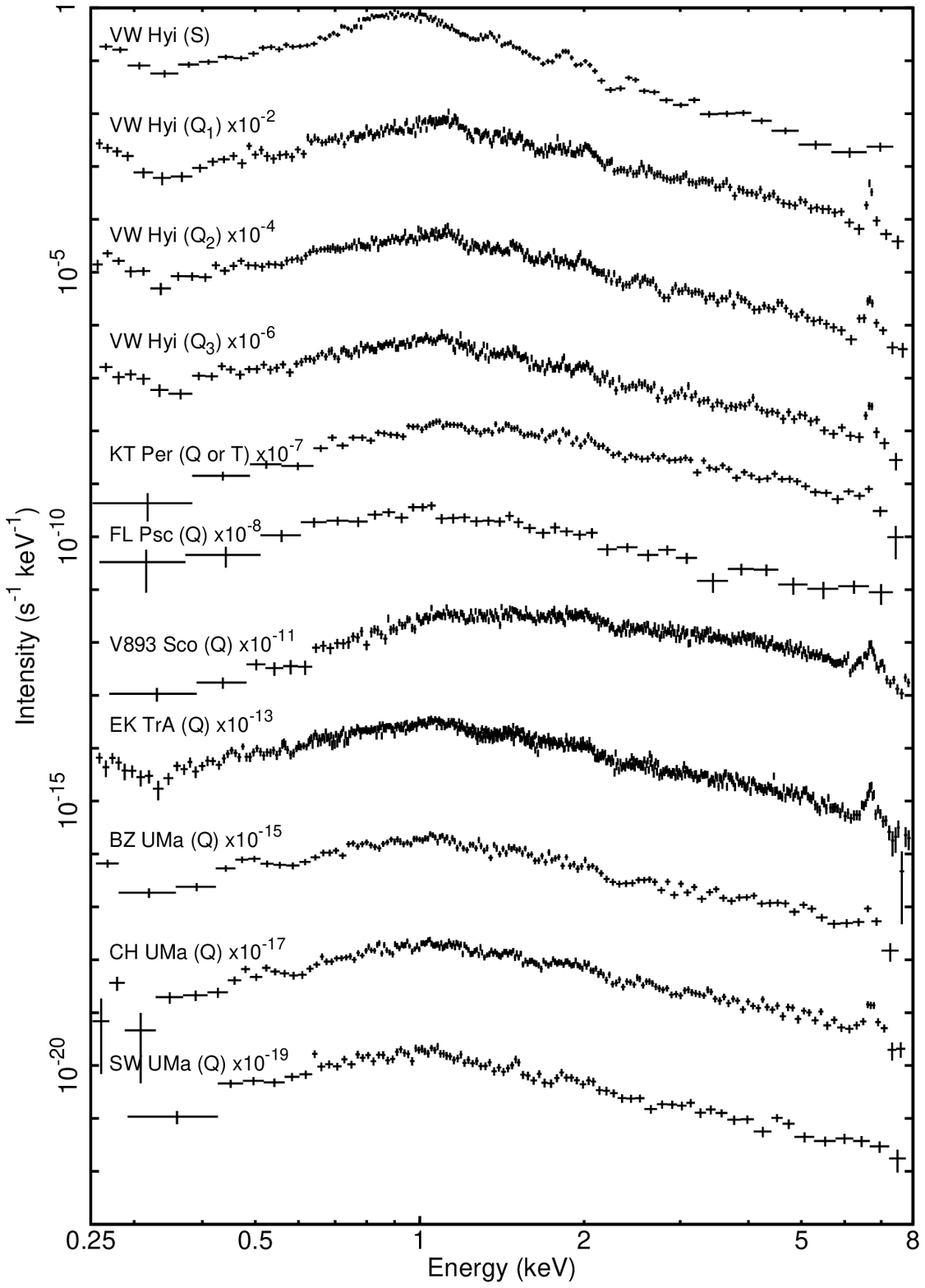}
  \end{center}
  \caption{{\it Continued.}}
\end{figure*}
%

\subsection{Spectral hardness}\label{s3-1}
\begin{figure}[ht]
  \begin{center}
    \includegraphics[width=80mm]{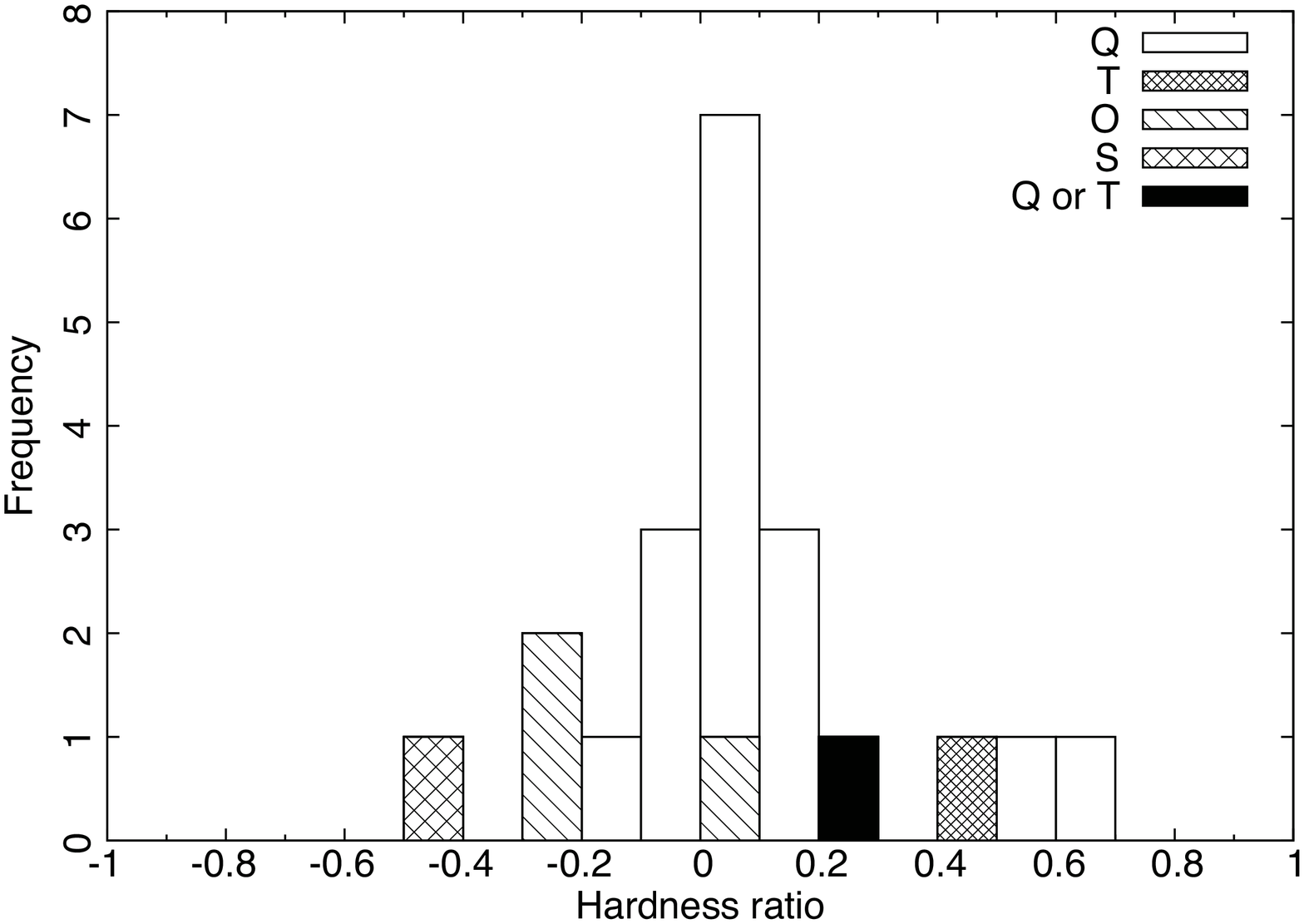}
  \end{center}
  \caption{Distribution of hardness ratio $(H-S)/(H+S)$. Different symbols show
 different states shown in the legend.}\label{f3}
\end{figure}

In order to have an overview of the spectral properties, we first investigated the
hardness ratio (HR) defined as $(H-S)/(H+S)$. Figure~\ref{f3} shows the histogram of
HR with different symbols for different states. The HR distribution shows that
different states have different distributions despite the inhomogeneity of the date
set: the transitional state exhibits hard spectra, and the outburst and the
super-outburst states exhibit softer spectra than those in the quiescent state. This 
indicates that the state is a primary factor to characterize X-ray spectra of DNe.

%

\subsection{Fiducial model}\label{s3-2}
For spectral fitting, we begin with the isobaric cooling flow model ({\tt mkcflow}:
\cite{mushotzky88}). In this model, the accreting gas cools as a steady flow with
gradually changing temperatures under an equal pressure, and each temperature layer
emits an optically-thin thermal plasma radiation. Despite its simplicity, the model
is known to describe the X-ray spectra of DNe very well at least in the quiescent
state (e.g., \cite{mukai03}; \cite{pandel05}). The free parameters are the maximum
and minimum temperatures of the plasma ($T_{\rm{max}}$ and $T_{\rm{min}}$), the metal
abundances ($Z$), and the normalization. The normalization represents the mass
accretion rate ($\dot{M}$). We used the abundance table by \citet{wilms00} and varied
the relative abundance collectively for all metals. We fixed $T_{\rm{min}}$ to 80.8~eV
unless otherwise noted. This is the lower limit value of the model in the XSPEC
fitting package, which we assume to be the temperature of the WD surface.

We fitted the background-subtracted spectra using the cooling flow model attenuated
by a photoelectric absorption ({\tt tbabs}: \cite{wilms00}) to account for the
extinction by the interstellar medium (ISM). Some sources show excess emission at
6.4~keV, which is presumably from the reprocessed Fe fluorescence emission. We added
a Gaussian model for this line. We refer this model as the ``fiducial model'' in the 
rest of the paper. 

The fitting result is shown in table~\ref{t2}. We fixed the $N_{\rm{H}}$ values for
Z~Cam, SS~Cyg, U~Gem and VW~Hyi based on ultraviolet spectral studies of individual
sources (\cite{baskill05}, \cite{mauche88}, \cite{long96}, and \cite{polidan90}). For
the others, the $N_{\rm{H}}$ value was a free parameter. The metal abundance was tied
for all the spectra of the same source in the same state. We did not tie the metal
abundance for spectra of the same source in different states. This is because, in
general, the plasma temperature differs for different states, and the elements with
the most prominent line emission change. Thus, it is not unexpected to have different
best-fit values of metal abundance when they are thawed collectively.

It is remarkable that the result is quite distinctive between the quiescent state
and the other states. If we adopt the reduced $\chi^{2}$ value of 1.33 to be the
success or failure criterion of the fitting, all but two (BV~Cen and V893 Sco)
sources were fitted successfully with the fiducial model for the quiescent state.
The fiducial model is not successful to explain the spectra in the other states. 

\begin{landscape}
\begin{table*}[ht]
  \begin{center}
    \hspace*{-160mm} 
    \begin{minipage}{1.0\hsize}
    \caption{Best-fit parameters with the fiducial model in the 0.25--10.0~keV band.\footnotemark[$*$]}
    \label{t2}
    \begin{tabular}{lccccccccc}
      \hline
      \hline
      Target & State & Seq num & $N_{\rm H}$ & $T_{\rm max}$ & $Z$ & $\dot{M}$ & EW & $L_{\rm X}$ & $\chi^{2}_{\rm red}$ (dof) \\
         & & & ($\times$10$^{20}$~cm$^{-2}$) & (keV) & ($Z_{\solar}$)& ($\times$10$^{-11}$~$M_{\solar}$~yr$^{-1}$) & (eV) & ($\times$10$^{30}$~erg~s$^{-1}$) & \\
      \hline
      VY Aqr   & Q & 402043010 & $<$~3.3 & 18.4$^{+2.3}_{-1.2}$ & 1.0$^{+0.3}_{-0.2}$ & 0.070$^{+0.004}_{-0.006}$ & 225$^{+61}_{-61}$& 2.18$^{+0.03}_{-0.03}$ & 1.07 (189) \\
      SS Aur   & Q & 402045010 & 1.3$^{+0.7}_{-0.9}$ & 26.5$^{+2.1}_{-1.2}$ & 2.3$^{+0.3}_{-0.3}$ & 0.53$^{+0.02}_{-0.02}$ & 96$^{+27}_{-29}$ & 22.2$^{+0.2}_{-0.2}$ & 1.12 (316) \\
      Z Cam    & T & 404022010 & 0.4\footnotemark[$\dagger$] & 80$^{+4}_{-4}$ & 3.9$^{+0.2}_{-0.2}$ & 1.05$^{+0.03}_{-0.03}$ & 109$^{+14}_{-14}$ & 100.2$^{+0.3}_{-0.3}$ & 3.42 (2757) \\
               & O & 407016010 & 0.4\footnotemark[$\dagger$] & 3.8$^{+0.2}_{-0.1}$ & 0.5$^{+0.1}_{-0.1}$ & 0.39$^{+0.02}_{-0.02}$ & 996$^{+235}_{-252}$ & 2.47$^{+0.04}_{-0.04}$ & 1.96 (183) \\
      BV Cen   & Q & 407047010 & 51$^{+1}_{-1}$ & $<$~77.0 & 2.6$^{+0.2}_{-0.2}$ & 3.74$^{+0.07}_{-0.04}$ & 60$^{+20}_{-20}$ & 352.2$^{+1.5}_{-1.5}$ & 1.48 (1334) \\
      SS Cyg   & Q & 400006010 & 0.35\footnotemark[$\dagger$] & 52.5$^{+1.1}_{-0.7}$ & 1.22$^{+0.04}_{-0.04}$ & 1.50$^{+0.01}_{-0.02}$ & 65$^{+3}_{-7}$ & 103.5$^{+0.2}_{-0.2}$ & 1.13 (4200) \\
               & O & 400007010 & 0.35\footnotemark[$\dagger$] & 7.26$^{+0.09}_{-0.09}$ & 0.63$^{+0.02}_{-0.02}$ & 2.84$^{+0.06}_{-0.06}$ & 137$^{+27}_{-27}$ & 36.6$^{+0.1}_{-0.1}$ & 2.58 (1592) \\
      BF Eri   & Q & 407045010 & 1.3$^{+0.6}_{-0.7}$ & 10.2$^{+0.9}_{-0.4}$ & 0.04$^{+0.04}_{-0.03}$ & 6.84$^{+0.27}_{-0.48}$ & 144$^{+64}_{-70}$ & 118.7$^{+1.3}_{-1.3}$ & 1.02 (302) \\
      U Gem    & Q & 407034010 & 0.31\footnotemark[$\dagger$] & 26.2$^{+1.0}_{-0.6}$ & 2.0$^{+0.1}_{-0.1}$ & 0.224$^{+0.003}_{-0.005}$ & 53$^{+13}_{-14}$ & 9.21$^{+0.05}_{-0.05}$ & 1.07 (990) \\
               & O & 407035010 & 0.31\footnotemark[$\dagger$] & 16.5$^{+0.4}_{-0.3}$ & 1.7$^{+0.1}_{-0.1}$ & 0.69$^{+0.01}_{-0.01}$ & 139$^{+13}_{-13}$ & 19.7$^{+0.1}_{-0.1}$ & 1.60 (1161) \\
      VW Hyi   & S & 406009010 & 0.006\footnotemark[$\dagger$] & 2.18$^{+0.02}_{-0.03}$ & 0.53$^{+0.02}_{-0.03}$ & 1.54$^{+0.03}_{-0.02}$ & 694$^{+110}_{-108}$ & 4.77$^{+0.03}_{-0.03}$ & 1.78 (735) \\
               & Q & 406009020 & 0.006\footnotemark[$\dagger$] & 9.3$^{+0.1}_{-0.1}$ & 1.80$^{+0.07}_{-0.06}$ & 0.53$^{+0.01}_{-0.01}$ & 23$^{+15}_{-15}$ & 9.08$^{+0.05}_{-0.06}$ & 1.32 (664) \\
               & Q & 406009030 & 0.006\footnotemark[$\dagger$] & 10.0$^{+0.1}_{-0.1}$ & 1.80 (tied) & 0.42$^{+0.01}_{-0.01}$ & 22$^{+15}_{-15}$ & 7.54$^{+0.05}_{-0.05}$ & 1.15 (680) \\
               & Q & 406009040 & 0.006\footnotemark[$\dagger$] & 9.8$^{+0.1}_{-0.1}$ & 1.80 (tied) & 0.45$^{+0.01}_{-0.01}$ & 9$^{+15}_{-9}$ & 7.96$^{+0.05}_{-0.05}$ & 1.15 (609) \\
      KT Per   & Q or T & 403041010 & 27$^{+2}_{-1}$ & 14.5$^{+0.4}_{-0.5}$ & 0.6$^{+0.1}_{-0.1}$ & 0.72$^{+0.03}_{-0.02}$ & 62$^{+25}_{-25}$ & 17.9$^{+0.1}_{-0.2}$ & 1.06 (417) \\
      FL Psc   & Q & 403039010 & $<$~1.0 & 15.0$^{+1.7}_{-2.5}$ & 1.2$^{+0.4}_{-0.5}$ & 0.048$^{+0.005}_{-0.003}$ & 228$^{+130}_{-130}$ & 1.25$^{+0.03}_{-0.03}$ & 1.00 (127) \\
      V893 Sco & Q & 401041010 & 85$^{+2}_{-1}$ & 35.7$^{+0.7}_{-1.3}$ & 2.0$^{+0.1}_{-0.1}$ & 1.49$^{+0.04}_{-0.02}$ & 47$^{+10}_{-10}$ & 77.7$^{+0.4}_{-0.4}$ & 1.34 (1077) \\
      EK TrA   & Q & 407044010 & 2.5$^{+0.4}_{-0.3}$ & 12.4$^{+0.1}_{-0.2}$ & 1.2$^{+0.1}_{-0.1}$ & 1.17$^{+0.02}_{-0.01}$ & 24$^{+10}_{-10}$ & 25.7$^{+0.1}_{-0.1}$ & 1.12 (1593) \\
      BZ UMa   & Q & 402046010 & $<$~0.21 & 13.7$^{+0.6}_{-0.4}$ & 1.0$^{+0.1}_{-0.1}$ & 1.07$^{+0.02}_{-0.03}$ & 59$^{+25}_{-25}$ & 25.7$^{+0.2}_{-0.2}$ & 1.02 (471) \\
      CH UMa   & Q & 407043010 & 4.9$^{+0.5}_{-0.6}$ & 15.0$^{+0.5}_{-0.3}$ & 1.4$^{+0.1}_{-0.1}$ & 5.4$^{+0.1}_{-0.2}$ & 68$^{+17}_{-17}$ & 140.2$^{+0.8}_{-0.8}$ & 1.18 (761) \\
      SW UMa   & Q & 402044010 & 0.4$^{+0.8}_{-0.4}$ & 7.5$^{+0.4}_{-0.3}$ & 0.6$^{+0.1}_{-0.1}$ & 0.69$^{+0.03}_{-0.03}$ & 197$^{+65}_{-71}$ & 9.4$^{+0.1}_{-0.1}$ & 1.15 (226) \\
      \hline
      \multicolumn{10}{l}{\parbox{180mm}{
          \footnotesize
          \par \noindent
          \footnotemark[$*$] Parameters are for the hydrogen column density
          ($N_{\rm{H}}$), the maximum plasma temperature ($T_{\rm{max}}$), the metal
          abundance with respect to the solar value ($Z$), the mass accretion rate
          ($\dot{M}$), the equivalent width of 6.4~keV line (EW) and the luminosity
          in the 0.5--10.0~keV band ($L_{\rm{X}}$). The reduced~$\chi^{2}$
          ($\chi^{2}_{\rm{red}}$) and the degree of freedom (dof) are also shown for
          the goodness of the fitting. The errors indicate a 1$\sigma$ statistical
          uncertainty. \\
          \footnotemark[$\dagger$] Hydrogen column density was fixed based on
     	  ultraviolet observations:
          Z~Cam;~\citet{baskill05}, SS~Cyg;~\citet{mauche88},
          U~Gem;~\citet{long96}, VW~Hyi;~\citet{polidan90}.
     }}
    \end{tabular}
    \end{minipage}
  \end{center}
\end{table*}
\end{landscape}

\subsection{Modified model by an extra extinction}\label{s3-3}
We inspected the spectra that were not fitted with the fiducial model, and found that
the fitting can be improved by several different modifications for different groups
of sources. 

The first modification is to add an extinction, presumably by the intrinsic absorber,
in addition to the ISM extinction. Three spectra (V893~Sco and BV~Cen in the
quiescent state and Z~Cam in the transitional state) showed an improved fitting
result by this modification. An intrinsic extinction should be also
considered for sources with a spectrum fitted by the fiducial model if the best-fit
$N{_{\mathrm{H}}}$ value is too large for the ISM extinction. We found that KT~Per is 
such a source.

\begin{figure*}[ht]
  \begin{minipage}{0.5\hsize}
    \begin{center}
      \includegraphics[width=80mm]{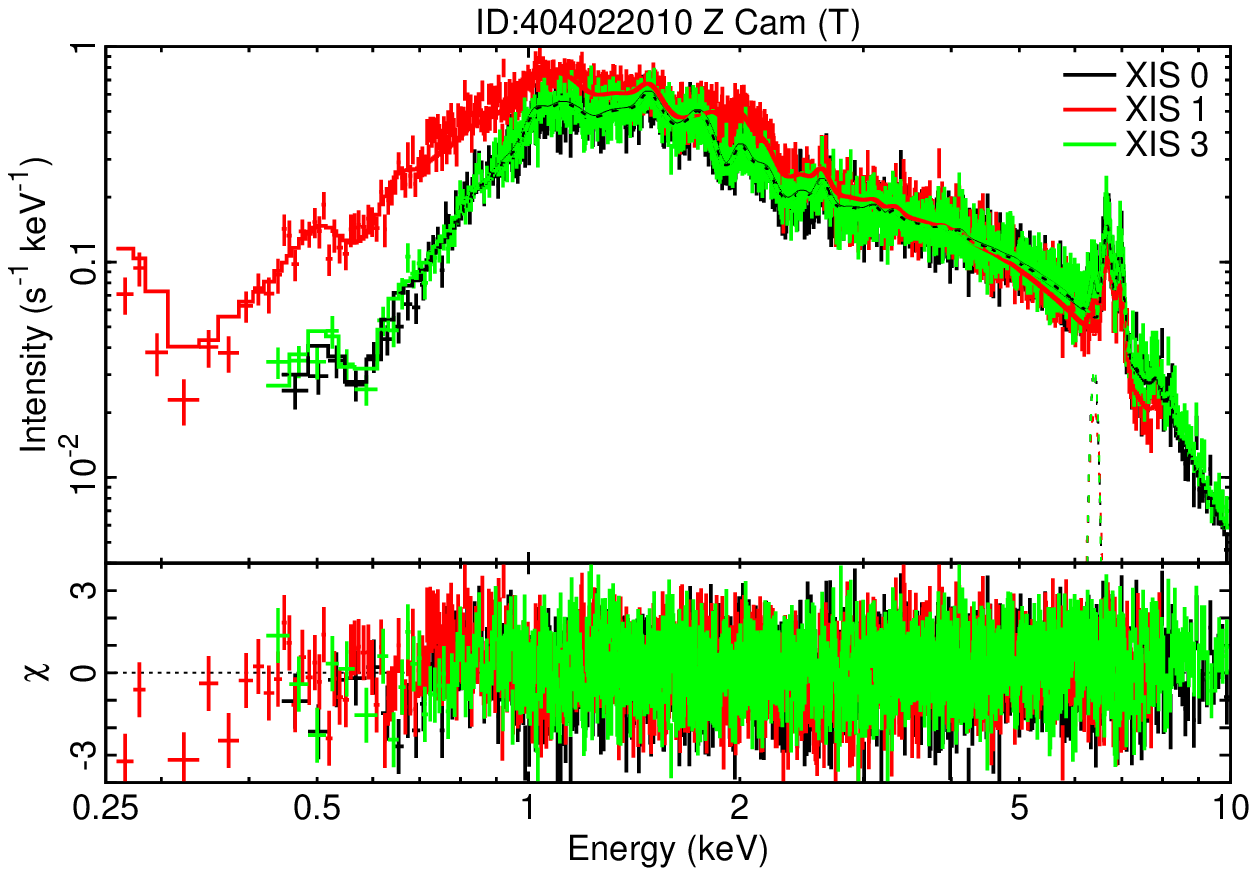}
    \end{center}
  \end{minipage}
  \begin{minipage}{0.5\hsize}
    \begin{center}
      \includegraphics[width=80mm]{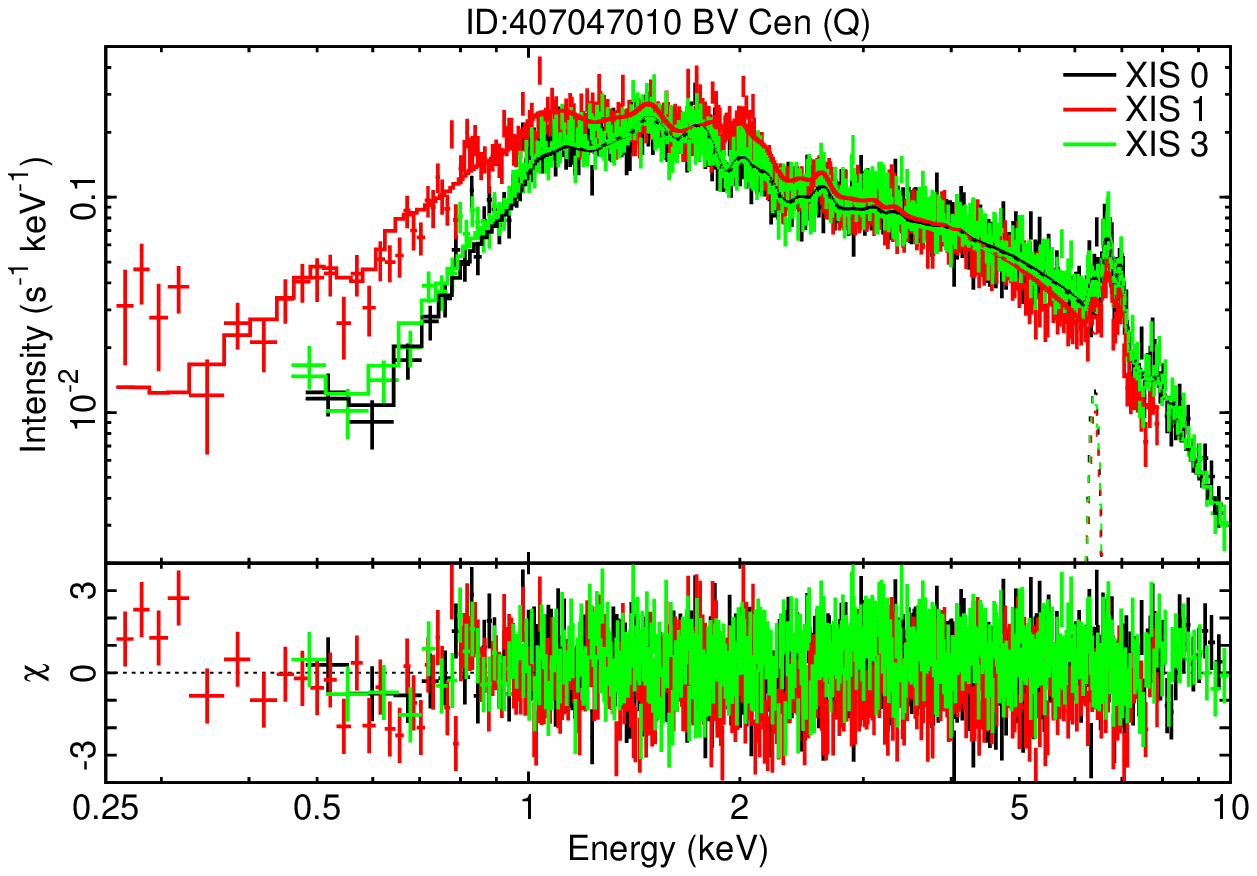}
    \end{center}
  \end{minipage}
  \begin{minipage}{0.5\hsize}
    \begin{center}
      \includegraphics[width=80mm]{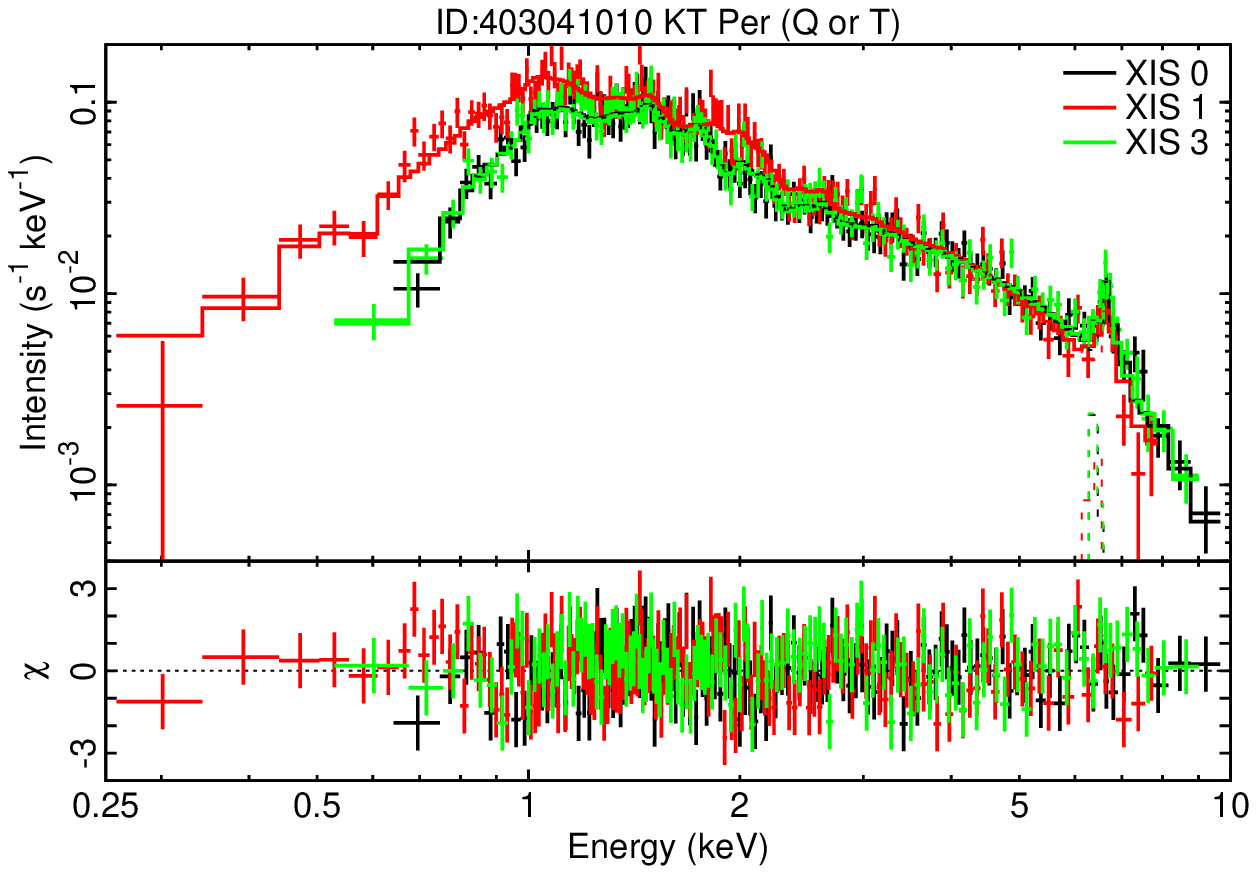}
    \end{center}
  \end{minipage}
  \begin{minipage}{0.5\hsize}
    \begin{center}
      \includegraphics[width=80mm]{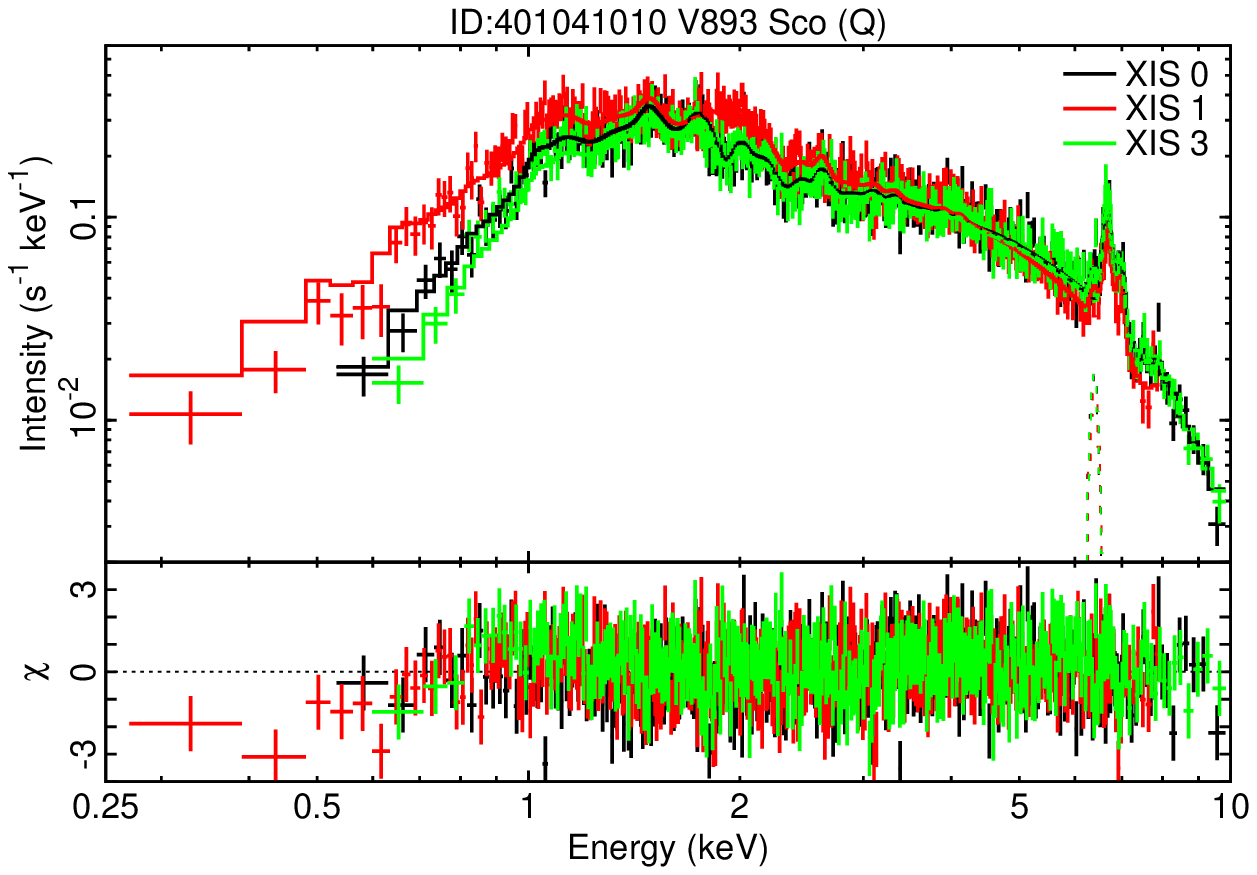}
    \end{center}
  \end{minipage}
  \vspace{1mm}
 \caption{Spectra and best-fit models (top) and the residuals (bottom) for the
   spectra fitted with the fiducial model modified by an additional neutral partial
   absorption component.}\label{f4}
\end{figure*}

For the four sources, we fixed the ISM extinction column to the value derived from
the B--V color (table~\ref{t3}) in the fiducial model. Then, we multiplied an
additional extinction model to represent the intrinsic absorber extinction. Two
models were employed: full neutral extinction ({\tt tbabs}) and a partial neutral
extinction ({\tt pcfabs}). The latter yielded a statistically better and acceptable
fit for all the spectra. The best-fit model and parameters with the partial
extinction model are shown in figure~\ref{f4} and table~\ref{t3}.

\begin{figure*}[ht]
  \begin{minipage}{0.5\hsize}
    \begin{center}
      \includegraphics[width=80mm]{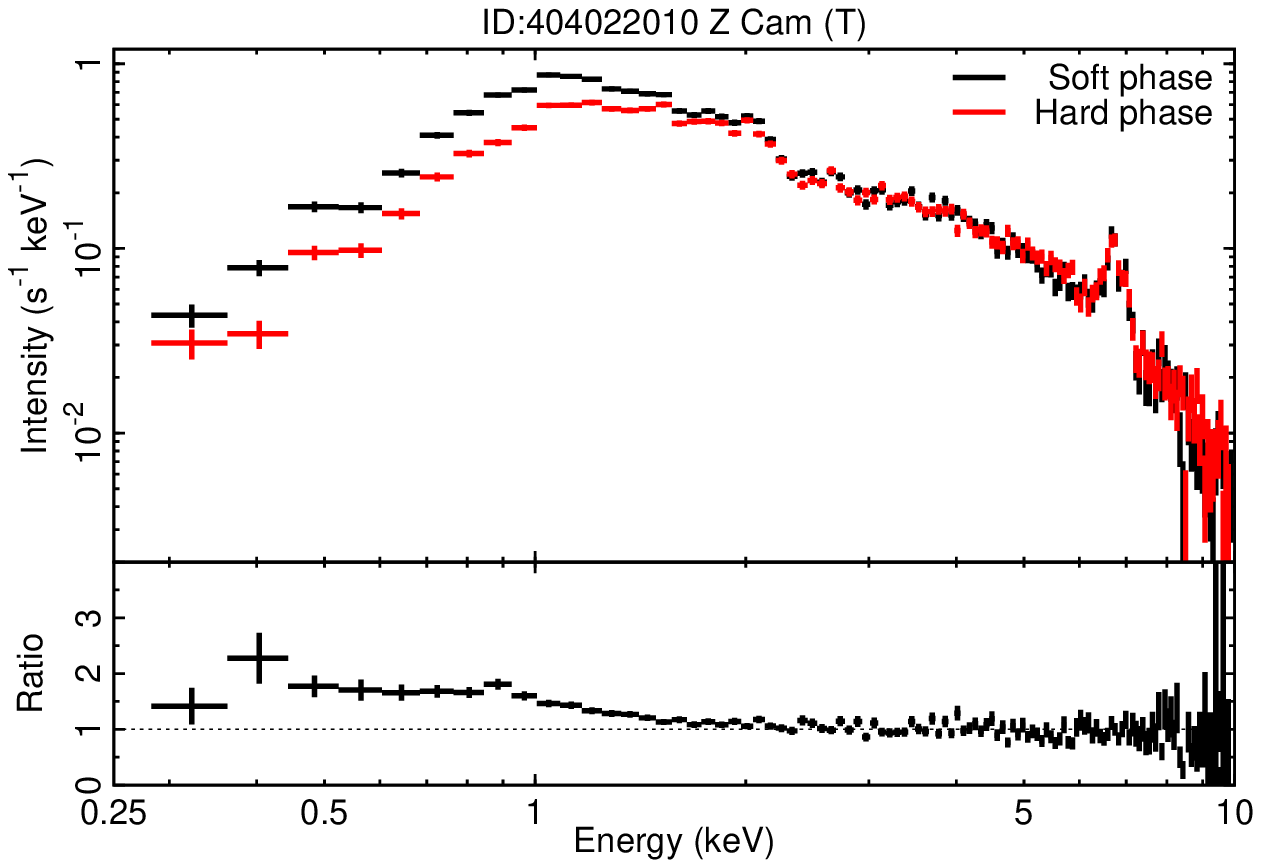}
    \end{center}
  \end{minipage}
  \begin{minipage}{0.5\hsize}
    \begin{center}
      \includegraphics[width=80mm]{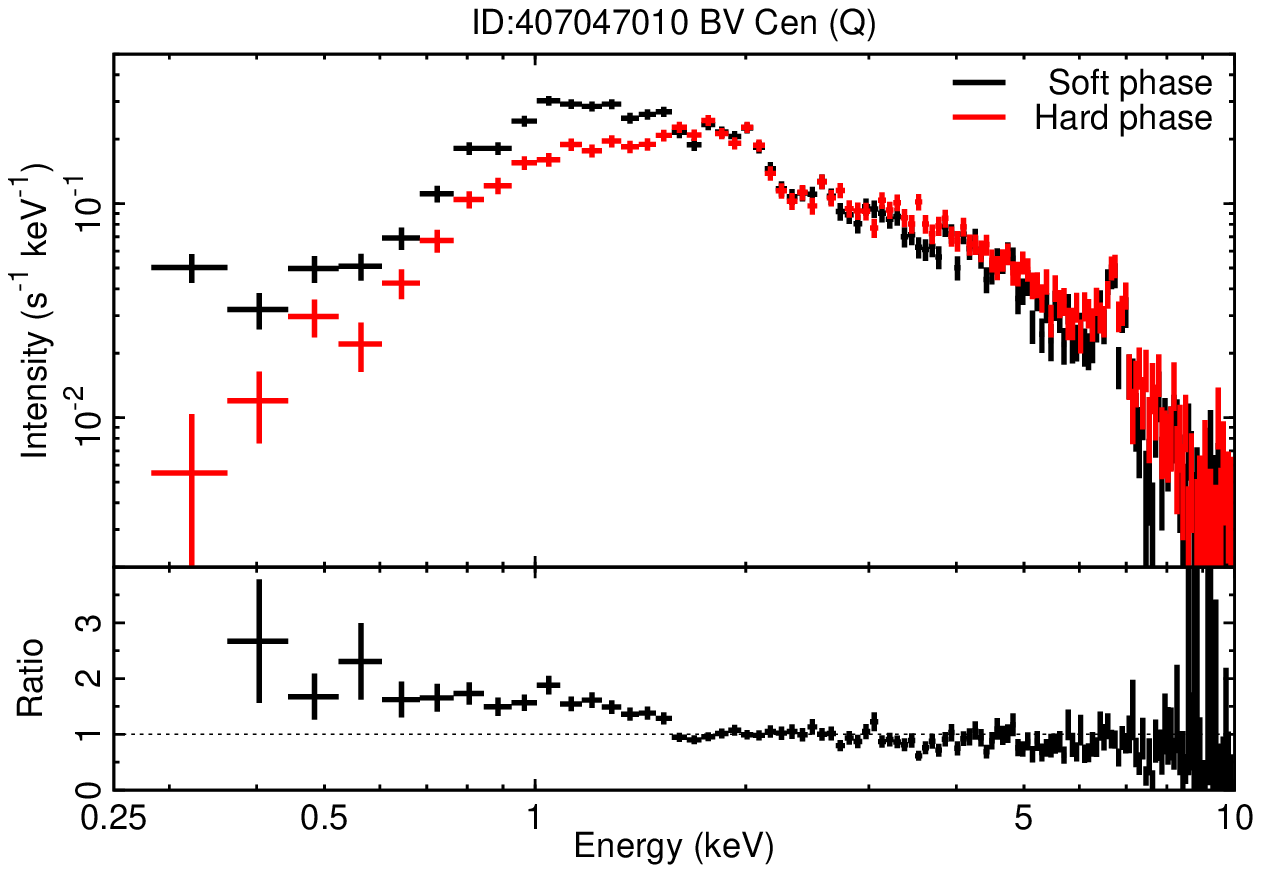}
    \end{center}
  \end{minipage}
  \begin{minipage}{0.5\hsize}
    \begin{center}
      \includegraphics[width=80mm]{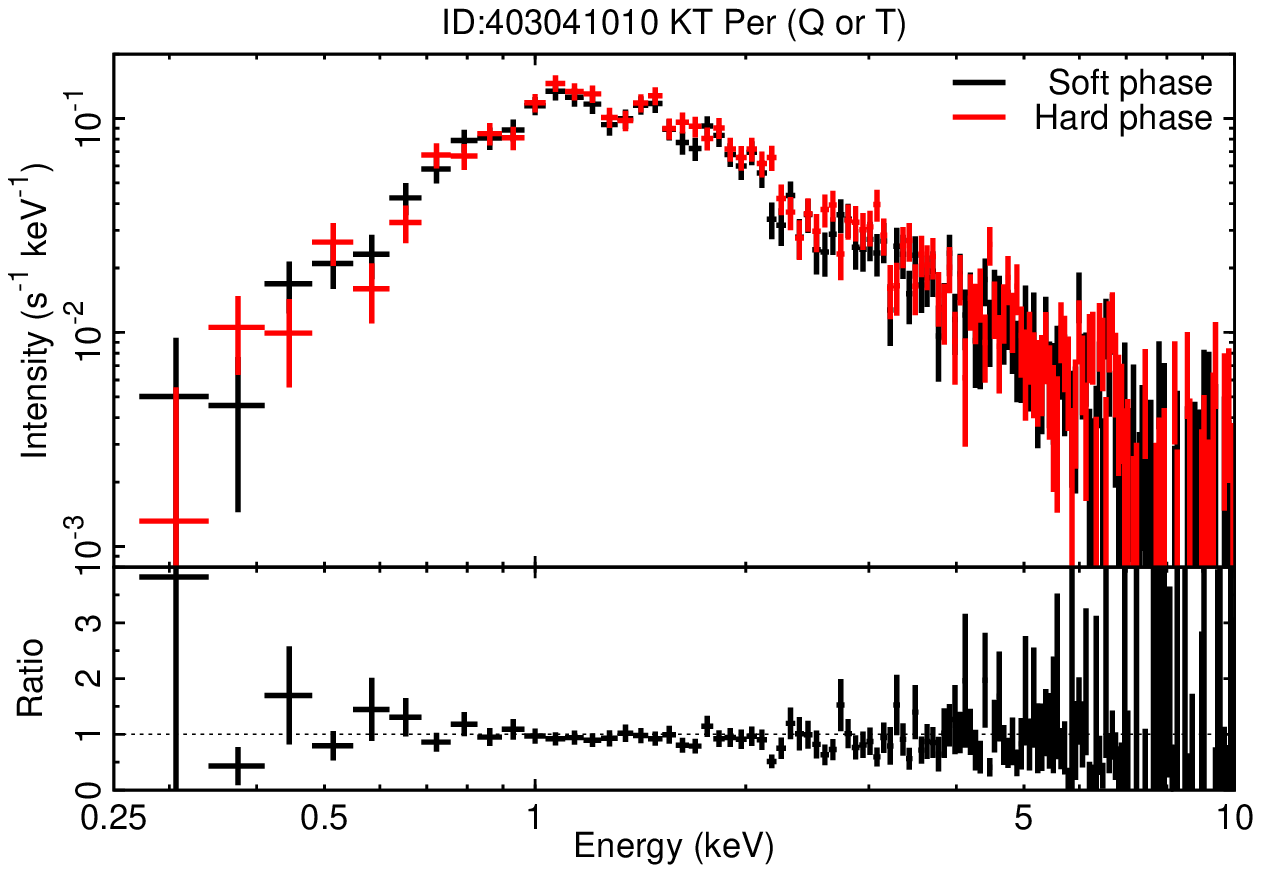}
    \end{center}
  \end{minipage}
  \begin{minipage}{0.5\hsize}
    \begin{center}
      \includegraphics[width=80mm]{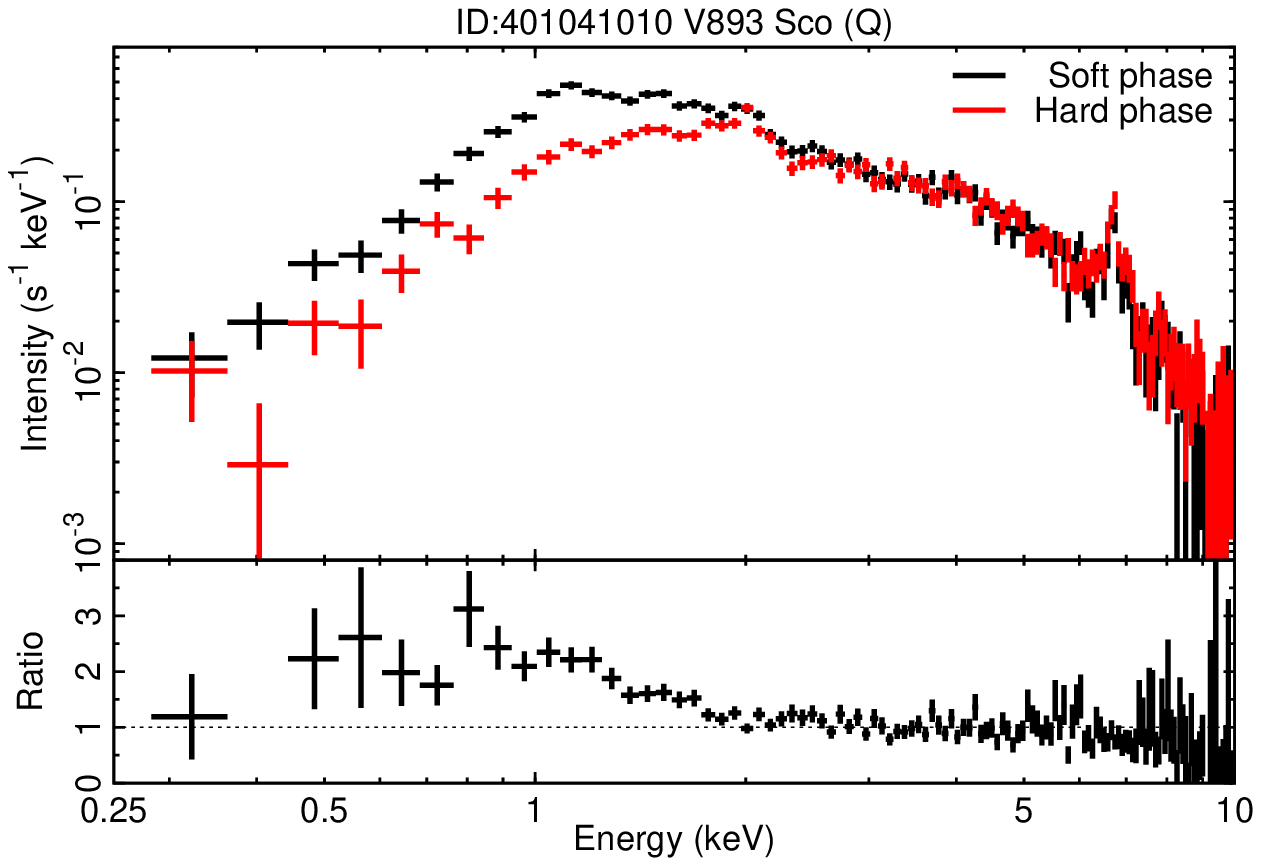}
    \end{center}
  \end{minipage}
  \vspace{1mm}
  \caption{XIS1 spectra in the soft (black) and hard (red) phase in the top panels
    and their ratio in the bottom panels.}\label{f5}
\end{figure*}

\begin{landscape}
\begin{table*}[ht]
  \renewcommand{\thefootnote}{\fnsymbol{footnote}}
  \begin{center}
    \hspace*{-160mm} 
    \begin{minipage}{1.0\hsize}
    \caption{Best-fit parameters for {\tt pcfabs} model.\footnotemark[$*$]}
    \label{t3}
    \begin{tabular}{lcccccccccccc}
      \hline
      \hline
      Target & State & Seq num & $N_{\rm H}^{\rm ISM}$\footnotemark[$\dagger$] & $N_{\rm H}^{\rm int.}$\footnotemark[$\ddagger$] & $C^{\rm int.}$\footnotemark[$\ddagger$] & $T_{\rm{max}}$ & $Z$ & $\dot{M}$ & EW & $L_{\rm X}$ & $\chi^{2}_{\rm red}$ (dof) \\
      & & & ($\times$10$^{20}$~cm$^{-2}$) & ($\times$10$^{22}$~cm$^{-2}$) & & (keV) & ($Z_{\solar}$) & ($\times$10$^{-11}$~$M_{\solar}$~yr$^{-1}$)& (eV) & ($\times$10$^{30}$~erg~s$^{-1}$)  \\
      \hline
      Z Cam    & T & 404022010 & 0.4 & 1.07$^{+0.03}_{-0.03}$ & 0.67$^{+0.01}_{-0.01}$ & 27.6$^{+0.3}_{-0.5}$ & 2.2$^{+0.1}_{-0.1}$ & 3.04$^{+0.05}_{-0.04}$ & 65$^{+5}_{-5}$ & 130.3$^{+0.3}_{-0.3}$ & 1.16 (2755) \\
      BV Cen   & Q & 407047010 & 5.8 & 1.42$^{+0.05}_{-0.04}$ & 0.78$^{+0.01}_{-0.01}$ & 27.5$^{+0.7}_{-0.7}$ & 1.9$^{+0.1}_{-0.1}$ & 9.21$^{+0.27}_{-0.23}$ & 65$^{+8}_{-10}$ & 394.4$^{+1.7}_{-1.7}$ & 1.20 (1333) \\
      KT Per & Q or T & 403041010 & 3.9 & 0.72$^{+0.10}_{-0.08}$ & 0.69$^{+0.03}_{-0.03}$ & 10.5$^{+0.8}_{-0.7}$ & 0.7$^{+0.1}_{-0.1}$ & 1.04$^{+0.07}_{-0.08}$ & 73$^{+23}_{-25}$ & 19.2$^{+0.2}_{-0.2}$ & 0.99 (416) \\
      V893 Sco & Q & 401041010 & 2.3 & 1.58$^{+0.06}_{-0.05}$ & 0.88$^{+0.01}_{-0.01}$ & 19.2$^{+0.6}_{-0.6}$ & 1.7$^{+0.1}_{-0.1}$ & 2.69$^{+0.09}_{-0.09}$ & 52$^{+6}_{-13}$ & 86.6$^{+0.4}_{-0.4}$ & 1.02 (1076) \\
      \hline
      \multicolumn{12}{l}{\parbox{220mm}{
          \footnotesize
          \par \noindent
          \footnotemark[$*$] The errors indicate a 1$\sigma$ statistical
          uncertainty. \\
          \footnotemark[$\dagger$] The hydrogen column density in ISM
          ($N_{\rm{H}}^{\rm{ISM}}$) are fixed to the value derived from $E(B-V)$
          measurements for Z~Cam \citep{baskill05}, BV~Cen \citep{godon12} and KT~Per
          and V893~Sco \citep{ozdonmez15}. \\
          \footnotemark[$\ddagger$] The hydrogen column density
          ($N_{\rm{H}}^{\rm{int.}}$) and the covering fraction ($C^{\rm{int.}}$) of the
          the additional neutral partial covering absorption . \\
      }}
    \end{tabular}
    \end{minipage}
  \end{center}
\end{table*}
\end{landscape}
Next, we divided the events into two phases depending on the hardness ratio being
greater or smaller than the median value in order to study origin of the spectral
variations. We generated spectrum of each phase (figure~\ref{f5}). In all the
spectra, there are no spectral changes in the hard band. However, in the soft band,
the intensity is smaller during the hard phase. We fitted the total-band spectra in
the soft and hard phases separately with the modified model, and found that the
spectral variation is primarily explained by change of the partial covering fraction 
of the intrinsic extinction component (table~\ref{t4}).
\begin{table*}[ht]
  \renewcommand{\thefootnote}{\fnsymbol{footnote}}
  \begin{center}
    \begin{minipage}{1.0\hsize}
    \caption{Best-fit parameters for divided spectra.\footnotemark[$*$]}
    \label{t4}
    \begin{tabular}{lcccccccccc}
      \hline
      \hline
      Target & State & Seq num & Phase &  $N_{\rm H}^{\rm int.}$\footnotemark[$\dagger$] & $C^{\rm int.}$\footnotemark[$\dagger$] & $T_{\rm{max}}$ & $\dot{M}$ & EW & $\chi^{2}_{\rm red}$ (dof) \\
      & & & & ($\times$10$^{22}$~cm$^{-2}$) & & (keV)  & ($\times$10$^{-11}$~$M_{\solar}$~yr$^{-1}$)& (eV) \\
      \hline
      Z Cam    & T & 404022010 & Hard & 0.80$^{+0.06}_{-0.04}$ & 0.58$^{+0.01}_{-0.01}$ & 28.5$^{+0.8}_{-1.1}$ & 2.84$^{+0.09}_{-0.06}$ & 62$^{+8}_{-8}$ & 1.11 (1657) \\
               &   &           & Soft & 1.20$^{+0.04}_{-0.03}$ & 0.79$^{+0.01}_{-0.01}$ & 28.1$^{+0.6}_{-0.7}$ & 3.12$^{+0.09}_{-0.07}$ & 70$^{+8}_{-8}$ & 1.14 (1514) \\
      BV Cen   & Q & 407047010 & Hard & 1.10$^{+0.08}_{-0.07}$ & 0.70$^{+0.01}_{-0.01}$ & 27.2$^{+1.5}_{-1.8}$ & 8.5$^{+0.4}_{-0.3}$ & 67$^{+13}_{-14}$ & 1.21 (715) \\
               &   &           & Soft & 1.70$^{+0.06}_{-0.06}$ & 0.86$^{+0.01}_{-0.01}$ & 27.2$^{+1.2}_{-1.0}$ & 10.1$^{+0.4}_{-0.4}$ & 58$^{+15}_{-10}$ & 1.18 (704) \\
      KT Per & Q or T & 403041010 & Hard & 0.42$^{+0.11}_{-0.10}$ & 0.65$^{+0.93}_{-0.07}$ & 9.5$^{+0.6}_{-0.7}$ & 0.98$^{+0.06}_{-0.06}$ & 114$^{+63}_{-63}$ & 1.14 (194) \\
             &        &           & Soft & 0.89$^{+0.13}_{-0.12}$ & 0.81$^{+0.03}_{-0.02}$ & 11.3$^{+1.1}_{-1.1}$ & 1.15$^{+0.09}_{-0.09}$ & 61$^{+36}_{-35}$ & 1.16 (210) \\
      V893 Sco & Q & 401041010 & Hard & 1.19$^{+0.07}_{-0.05}$ & 0.85$^{+0.01}_{-0.01}$ & 20.9$^{+0.7}_{-0.8}$ & 2.43$^{+0.11}_{-0.09}$ & 40$^{+13}_{-13}$ & 0.98 (592) \\
               &   &           & Soft & 1.99$^{+0.08}_{-0.06}$ & 0.92$^{+0.01}_{-0.01}$ & 18.4$^{+0.7}_{-0.8}$ & 2.88$^{+0.14}_{-0.11}$ & 62$^{+13}_{-13}$ & 1.04 (519) \\
      \hline
      \multicolumn{10}{l}{\parbox{160mm}{
          \footnotesize
          \par \noindent
          \footnotemark[$*$] The errors indicate a 1$\sigma$ statistical uncertainty.
          The hydrogen column density in ISM ($N_{\rm{H}}^{\rm{ISM}}$) and the metal
          abundance ($Z$) are fixed to the best-fit value listed in table~\ref{t3}.\\
          \footnotemark[$\ddagger$] The hydrogen column density
          ($N_{\rm{H}}^{\rm{int.}}$) and the covering fraction ($C^{\rm{int.}}$) of the
          the additional neutral partial covering absorption . \\
      }}
    \end{tabular}
    \end{minipage}
  \end{center}
\end{table*}

\subsection{Modified model with extra soft-band emission}\label{s3-4}
For other spectra during the outburst or the super-outburst states, the fitting
with the fiducial model was not successful especially below 2~keV. This cannot be
explained by adding an extra extinction (\S~\ref{s3-3}). However, when we restricted
the energy range to 2--10~keV for the fiducial model fitting, the result was
successful; the $\chi^{2}_{\rm red}$ (and d.o.f) of Z~Cam, SS~Cyg, U~Gem, and VW Hyi
are 1.13 (55), 1.21 (582), 1.24 (508), and 1.25 (142), respectively. By extrapolating
the best-fit fiducial model to the range below 2~keV, we clearly see the excess with
line emission in the soft band, which suggests the presence of an additional
thin-thermal plasma component.

In the cooling flow component of our fiducial model, the power-law slope of the
differential emission measure distribution is fixed. \citet{pandel05} took a
different approach by treating the power-law slope as a free parameter. This allows a
more flexible distribution to be fitted. We followed this approach, but the soft
excess emission was still found, suggesting that the additional component is
something different from the BL plasma represented by a continuous differential
emission measure distribution.

It is also suggested that the black body emission from the BL emerges during outbursts
\citep{mauche95,long96}. The temperature is too low to constrain only from the XIS data,
so we refer to the values derived in EUVE studies: 20~eV for SS~Cyg \citep{mauche95},
11.9~eV for U~Gem \citep{long96}, and $\lesssim$10~eV for VW~Hyi \citep{mauche96}. Z~Cam
has no reference, but has no excess emission in the softest band of the XIS spectrum
like U~Gem, so we used 11.9~eV as an upper limit. We did spectral fitting with the black
body luminosity as an additional free parameter, and tested the fitting improvement by
the $F$ statistics. The addition of the black body component was only justified for SS
Cyg in outburst. We thus added a black body component only for this spectrum hereafter.

\begin{figure*}[htbp]
  \begin{minipage}{0.5\hsize}
    \begin{center}
      \includegraphics[width=80mm]{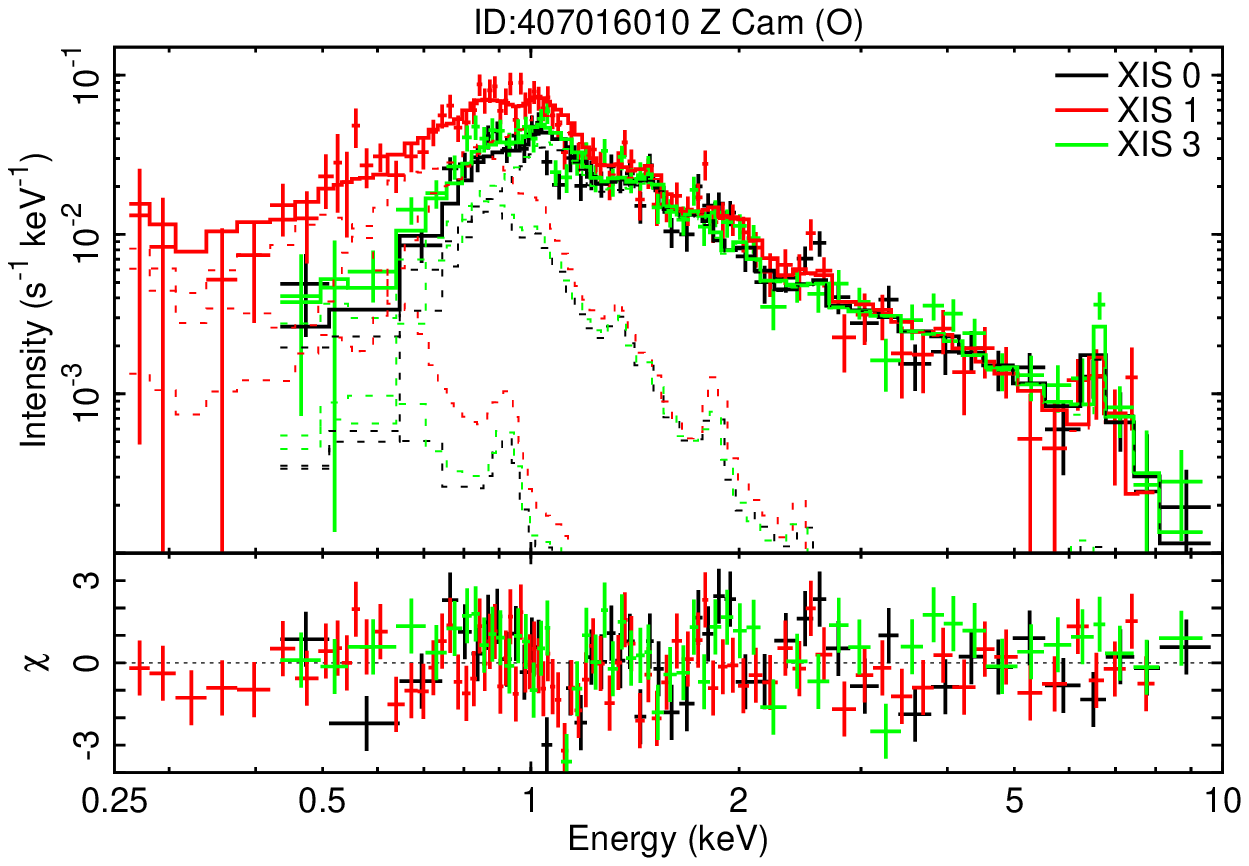}
    \end{center}
  \end{minipage}
  \begin{minipage}{0.5\hsize}
    \begin{center}
      \includegraphics[width=80mm]{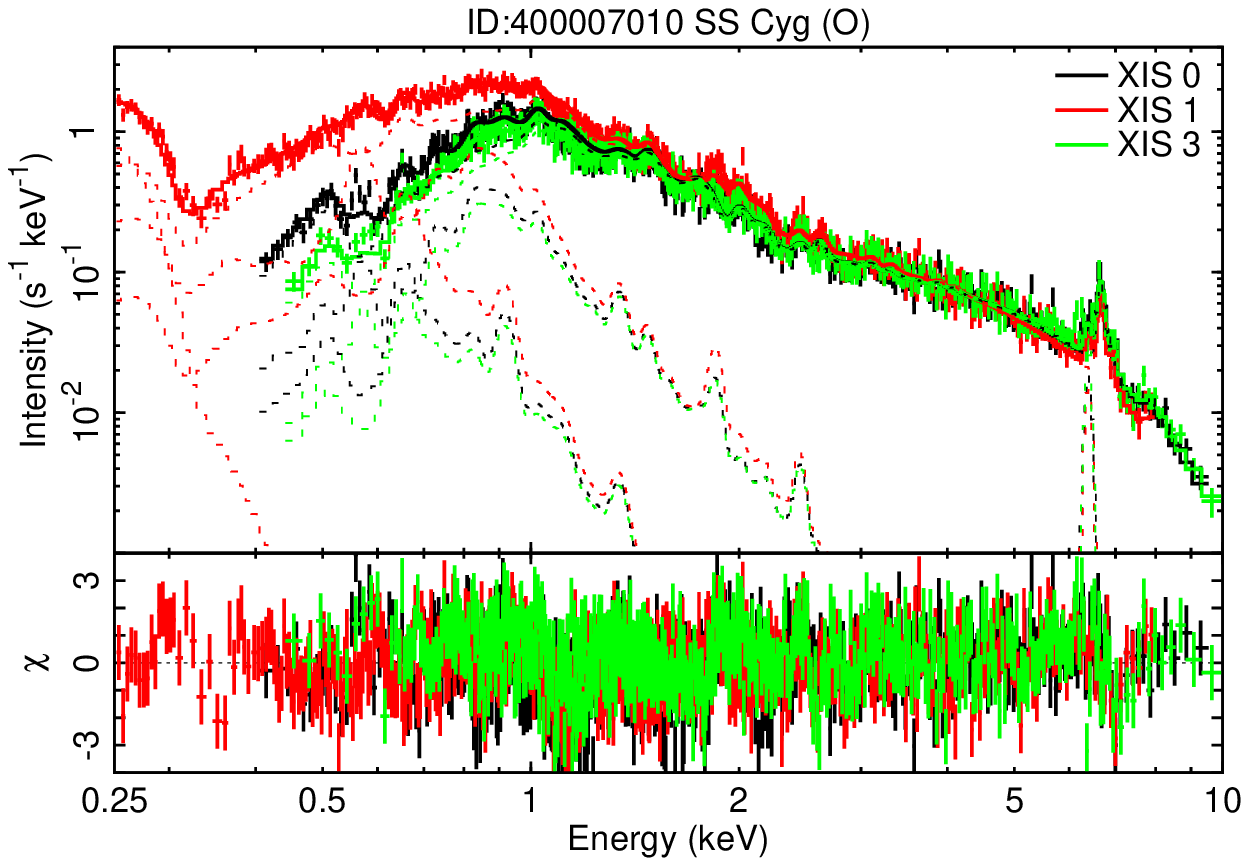}
    \end{center}
  \end{minipage}
  \begin{minipage}{0.5\hsize}
    \begin{center}
      \includegraphics[width=80mm]{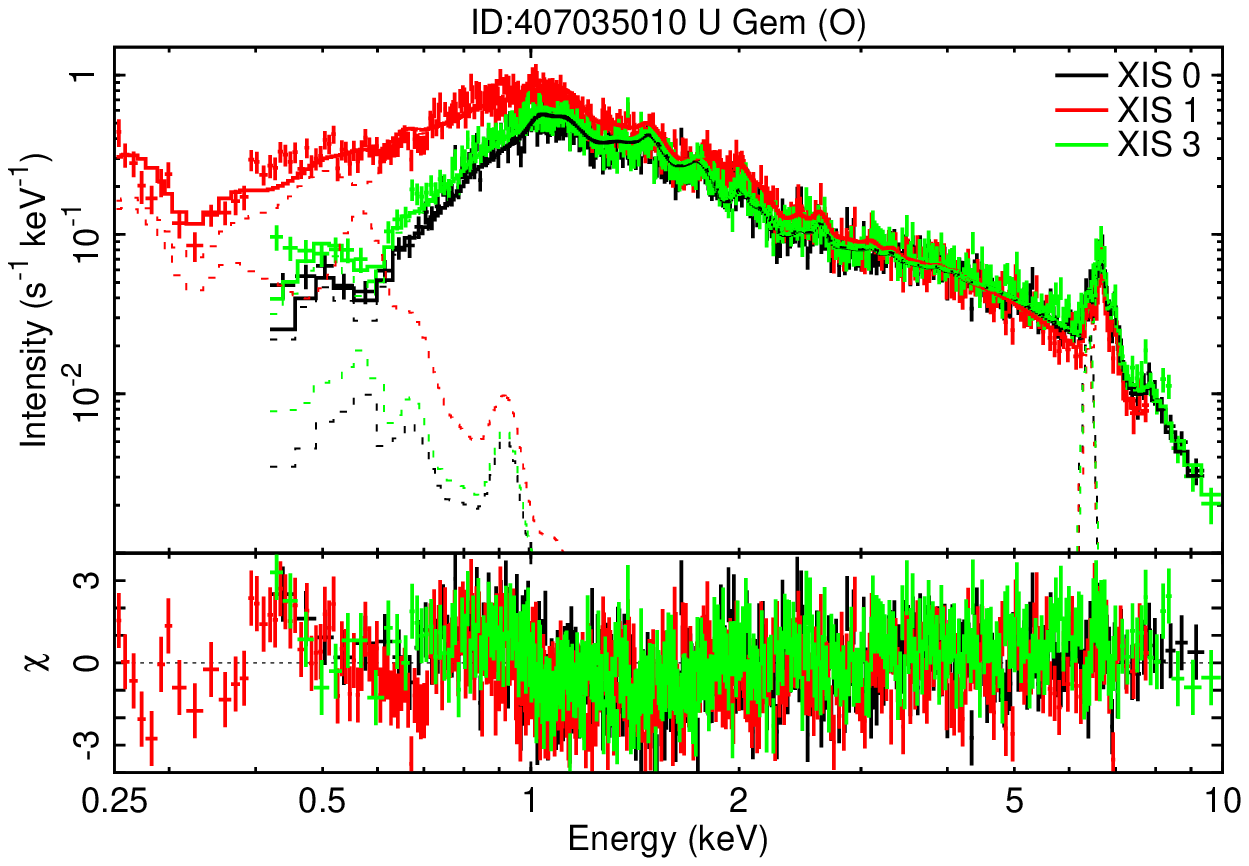}
    \end{center}
  \end{minipage}
  \begin{minipage}{0.5\hsize}
    \begin{center}
      \includegraphics[width=80mm]{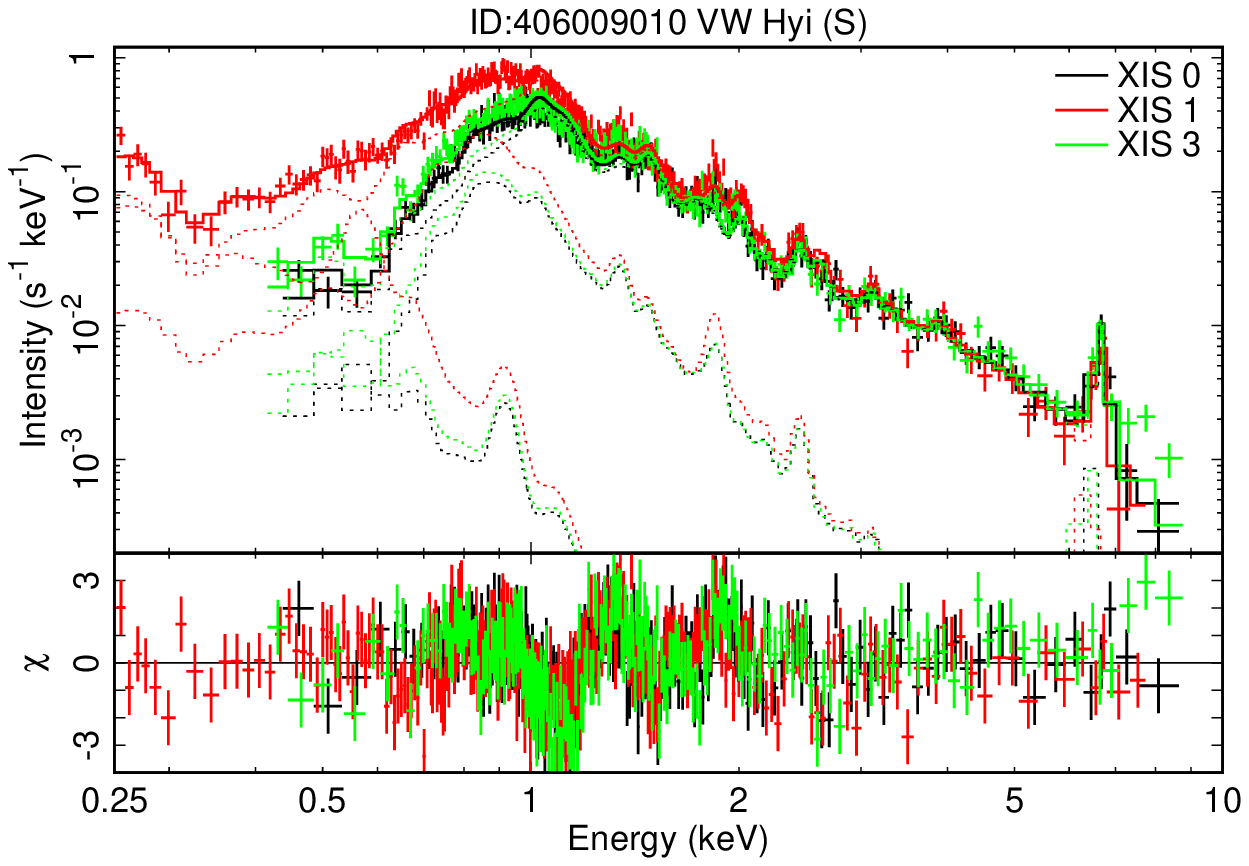}
    \end{center}
  \end{minipage}
  \vspace{1mm}
 \caption{Spectra and best-fit models (top) and the residuals (bottom) for the
   spectra fitted with the fiducial model modified by additional soft thermal
   emission.}\label{f6}
\end{figure*}

In the end, we evaluated this soft excess emission. After fixing the best-fit parameters
of the fiducial model in the 2--10~keV band, we added a single temperature thermal
plasma model ({\tt{MEKAL}}; \cite{mewe85, mewe86, liedahl95, kaastra96}). As a
consequence, the fitting improved by adding one thermal plasma component in U~Gem, or
two components are required in SS~Cyg and Z~Cam (table~\ref{t5} and
figure~\ref{f6}). The super-outburst of VW Hyi is, however, not explained even by adding
extra-soft components (table~\ref{t5} and figure~\ref{f6}).

\begin{landscape}
\begin{table*}[ht]
  \begin{center}
    \hspace*{-160mm} 
    \begin{minipage}{1.0\hsize}
    \caption{Best-fit parameters for outburst and super-outburst states in the
      0.25--10.0~keV band.\footnotemark[$*$]}
    \label{t5}
    \begin{tabular}{lccccccccccccccc}
      \hline
      \hline
      Target & State & Seq num & $N_{\rm H}$\footnotemark[$\dagger$] & $T_{\rm min}$ & $T_{\rm max}$ & $Z$\footnotemark[$\ddagger$] & $\dot{M}$ & $kT_{1}$ & $kT_{2}$ & $T_{\rm BB}$\footnotemark[$\S$] & EW & $L_{\rm X}$ & $\chi^{2}_{\rm red}$ (dof) \\
      & & & ($\times$10$^{20}$~cm$^{-2}$) & (keV) & (keV) & ($Z_{\solar}$)& ($\times$10$^{-11}$~$M_{\solar}$~yr$^{-1}$) & (keV) & (keV) & (eV) & (eV) & ($\times$10$^{30}$~erg~s$^{-1}$) \\
      \hline
      Z Cam  & O & 407016010 & 0.4 & 0.0808\footnotemark[$\P$] & 8.0$^{+1.6}_{-1.8}$ & 2.2 & 0.17$^{+0.06}_{-0.03}$ & 0.14$^{+0.02}_{-0.02}$ & 0.60$^{+0.03}_{-0.02}$ & --- & 36$^{+81}_{-36}$ & 2.68$^{+0.05}_{-0.04}$ & 1.57 (183) \\
      SS Cyg & O & 400007010 & 0.35 & 0.0808\footnotemark[$\P$] & 11.6$^{+0.5}_{-0.3}$ & 1.22 & 1.72$^{+0.05}_{-0.06}$ & 0.17$^{+0.01}_{-0.01}$ & 0.61$^{+0.01}_{-0.01}$ & 20.0 & 73$^{+10}_{-10}$ & 39.6$^{+0.1}_{-0.1}$ & 1.39 (1591) \\
      U Gem  & O & 407035010 & 0.31 & 0.0808\footnotemark[$\P$] & 15.6$^{+1.1}_{-0.5}$ & 2.0 & 0.74$^{+0.02}_{-0.04}$ & 0.12$^{+0.01}_{-0.01}$ & --- & --- & 125$^{+12}_{-13}$ & 19.9$^{+0.1}_{-0.1}$ & 1.51 (1163) \\
      VW Hyi & S & 406009010 & 0.006 & 0.0808\footnotemark[$\P$] & 3.28$^{+0.11}_{-0.14}$ & 1.80 & 0.72$^{+0.05}_{-0.04}$ & 0.12$^{+0.01}_{-0.01}$ & 0.62$^{+0.01}_{-0.01}$ & --- &  76$^{+37}_{-37}$ & 4.97$^{+0.03}_{-0.03}$ & 1.68 (735) \\
             &   &           & 0.006 & 1.20$^{+0.03}_{-0.03}$ & 3.28$^{+0.11}_{-0.14}$ & 1.80 & 0.72$^{+0.05}_{-0.04}$ & 0.13$^{+0.01}_{-0.02}$ & 0.66$^{+0.01}_{-0.01}$ & --- & 76$^{+37}_{-37}$ & 5.02$^{+0.03}_{-0.03}$ & 1.58 (734) \\
      \hline
      \multicolumn{11}{l}{\parbox{180mm}{
          \footnotesize
          \par \noindent
          \footnotemark[$*$] The errors indicate a 1$\sigma$ statistical
          uncertainty. $kT_{1}$ and ${kT_{2}}$ show the temperature of the additional
          thermal plasma component. \\
          \footnotemark[$\dagger$] The values of $N_{\rm H}$ were fixed. 
            (Z~Cam; \cite{baskill05}, SS~Cyg; \cite{mauche88},
            U~Gem; \cite{long96}, VW~Hyi; \cite{polidan90}) \\
          \footnotemark[$\ddagger$] Abundance is fixed to the best-fit values in the
          quiescent or the transitional state. \\
          \footnotemark[$\S$] The temperature of black body component for SS~Cyg. The
          component was not necessary for other spectra.\\
          \footnotemark[$\P$] $T_{\rm min}$ is fixed to 80.8~eV.\\
          }}
    \end{tabular}
  \end{minipage}
  \end{center}
\end{table*}
\end{landscape}

\subsection{Modified model further with $T_{\rm min}$}\label{s3-5}
\begin{figure}[ht]
  \begin{center}
    \includegraphics[width=80mm]{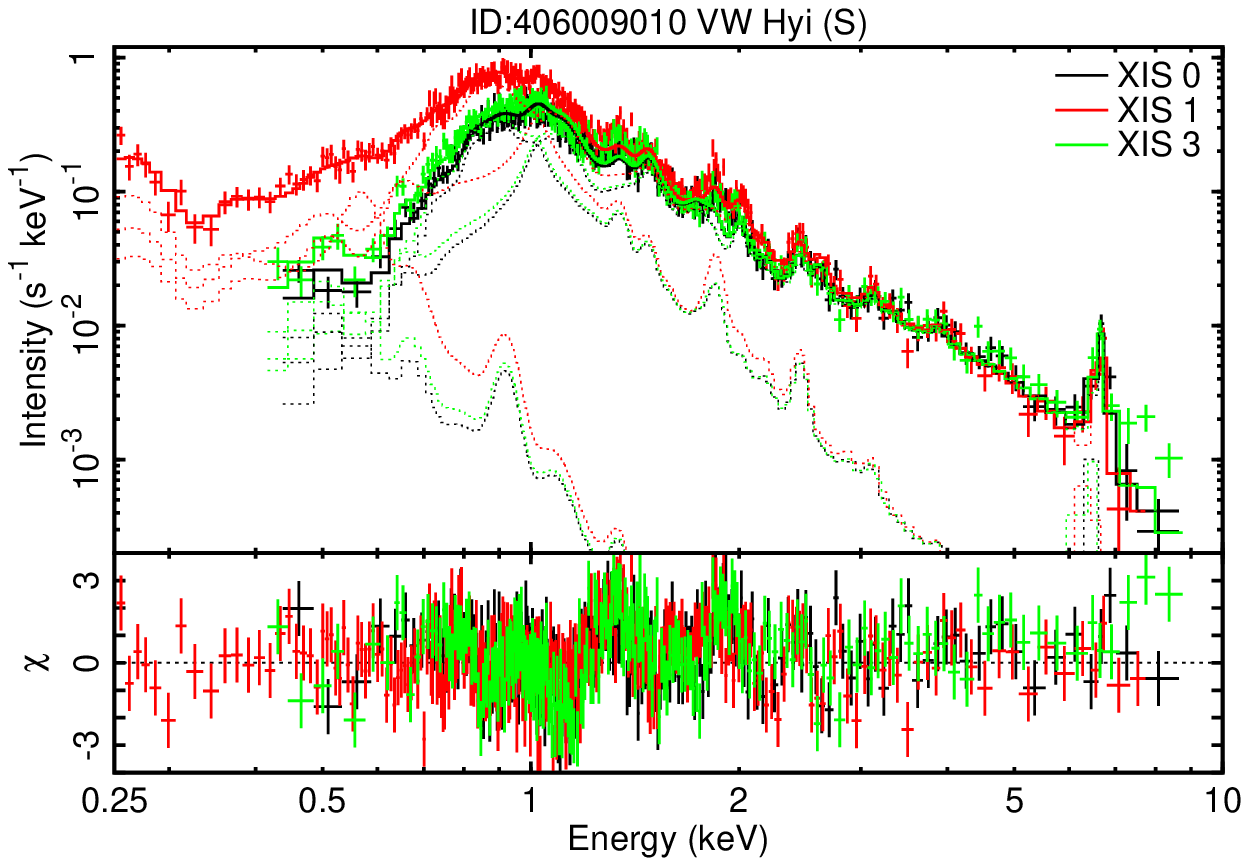}
  \end{center}
  \caption{Spectra and best-fit models (top) and the residuals (bottom) for the
    spectra fitted by altering the $T_{\rm min}$ value.}\label{f7}
\end{figure}

One spectrum remained without being fitted successfully by any modifications discussed
in \S~\ref{s3-3} and \S~\ref{s3-4}, is the VW~Hyi spectrum in the super-outburst
state. Since there are large residuals in the 0.9--1.2~keV band (figure~\ref{f6}), which
are presumably due to the Fe L emission line complex, we changed the value of $T_{\rm
min}$ in the cooling flow model to $\sim$1~keV, which is the only free parameter at this
point. Consequently, the fitting improved by changing $T_{\rm min}\sim$~1.20~keV with
the $F$ probability of $\sim$3.8$\times$10$^{-11}$ against the null hypothesis that the
improvement is random (table~\ref{t5} and figure~\ref{f6},~\ref{f7}).

\section{Discussion}\label{s4}
\subsection{Comparison with preceding results}\label{s4-1}
First, we compare our fitting results with those in preceding works using the same
data set. \citet{byckling10} selected 12 DNe with a parallax measurement among the
XMM-Newton, ASCA, and Suzaku archives for constructing a precise luminosity function
of DNe. Eight of the samples are the same with ours. Seven are in the quiescent
state, while one (KT Per) is in the outburst according to their classification. The
states are consistent with ours except for KT Per, which we classify as the quiescent
or transition state in our definition. They fitted all the spectra using the cooling
flow model similarly to our fiducial model, including KT Per. All the parameters are
consistent with ours except for the abundance. The difference of the abundance is
presumably due to the use of different abundance table; \citet{wilms00} used in this
work has about a half of Fe abundance than \citet{anders89} used in
\citet{byckling10}; Fe dominates the abundance determination in the fitting for
spectra in the quiescent state.

\citet{saitou12} and \citet{ishida09} studied individual sources (Z~Cam in outburst and
SS~Cyg in both quiescence and outburst, respectively) in detail. Both work used not only
the XIS but also the HXD data up to 40~keV, and fitted the spectra by considering the
reflection component. Since we do not include the reflection component, our $T_{\rm
max}$ and $\dot{M}$ are higher than their results. This also gives a slight change in
the estimate of the intrinsic absorber extinction model.  \citet{saitou12} derived a
larger covering fraction by 10\% and a smaller column density by 17\% than ours. For
other parameters, our results on Z~Cam and SS~Cyg are consistent with those by
\citet{saitou12} and \citet{ishida09}.

\citet{mukai09} studied V893 Sco and fitted the XIS spectrum based on the cooling
flow model. They showed that a partial covering extinction in addition to the ISM
extinction is needed to explain the XIS spectrum , which we confirmed.

\subsection{X-ray spectral characteristics of each state}\label{s4-2}
We started the spectral fitting with the fiducial model, and applied several
modifications finally to explain all the spectra. We took a data-oriented approach,
but the result shows that the best-fit model depends strongly on the state of DNe.
Despite the diversity of the samples belonging to different types, the result is
quite clear.

\begin{itemize}
  \item All but two (BV Cen and V893 Sco) spectra in the quiescent state were
        explained just by the fiducial model. The two exceptions require a partial
        covering absorption.
  \item The only spectrum definitively in the transitional state (Z Cam) and another
        possibly in the transitional state (KT Per) were explained by an additional
        partial covering extinction upon the fiducial model.
  \item All outburst spectra were explained by the fiducial model added by one or two
        thin-thermal plasma components in the soft band. 
  \item The only super-outburst spectrum required a further increase of $T_{\rm min}$.
\end{itemize}

In all the spectra, the fiducial model explained the spectra in the hard band, which
indicates that the hard X-ray emission arises from the BL in all the states.

\begin{figure*}[ht]
  \begin{minipage}{0.5\hsize}
    \begin{center}
      \includegraphics[width=80mm]{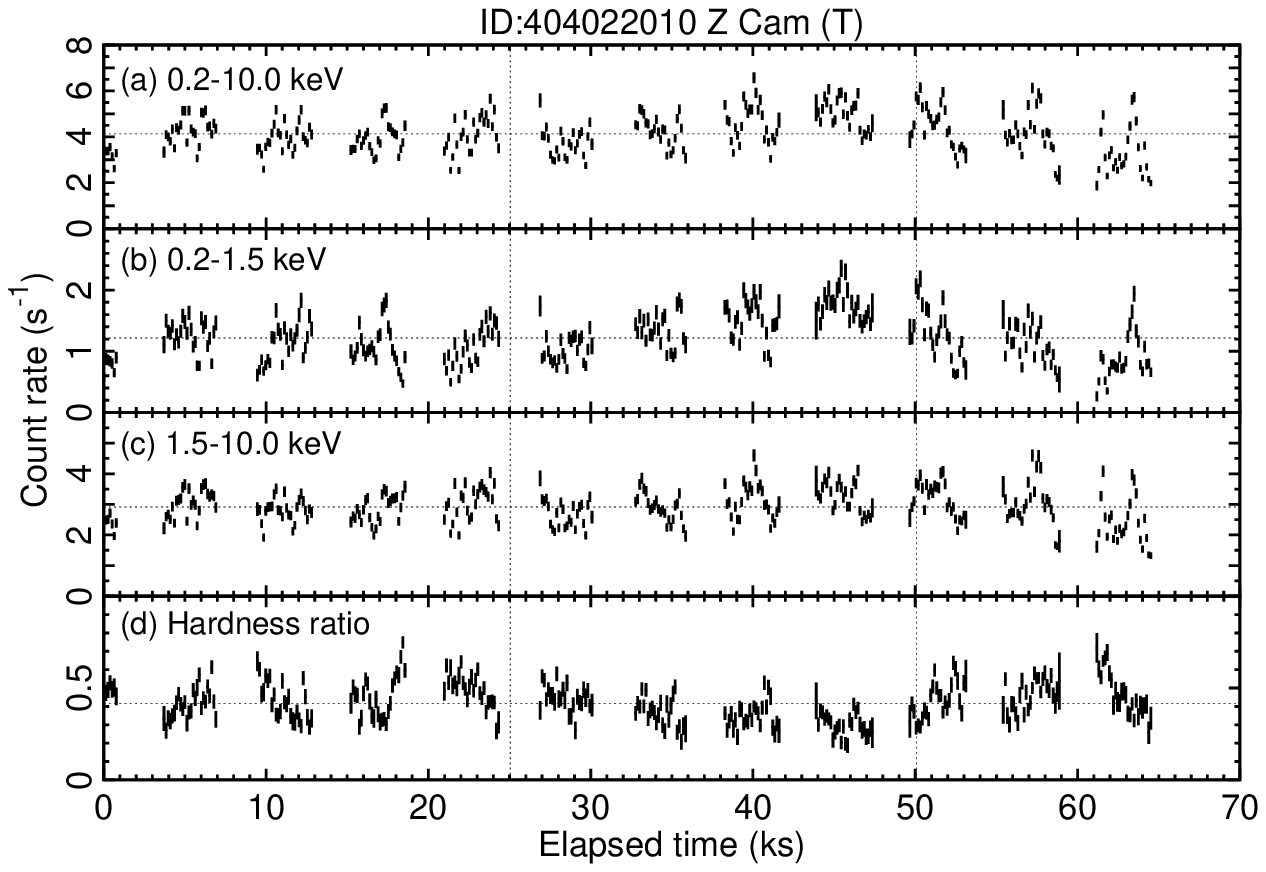}
    \end{center}
  \end{minipage}
  \begin{minipage}{0.5\hsize}
    \begin{center}
      \includegraphics[width=80mm]{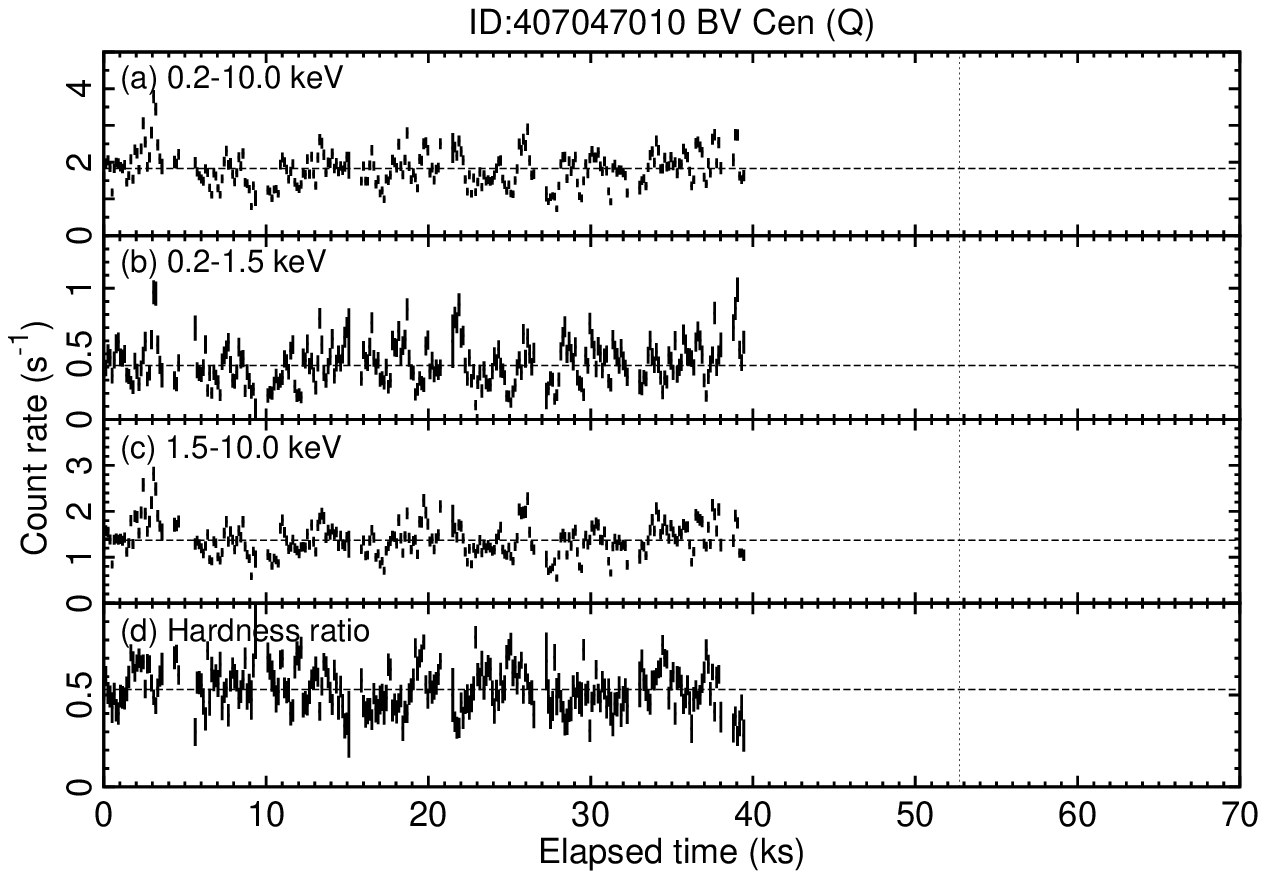}
    \end{center}
  \end{minipage}
  \begin{minipage}{0.5\hsize}
    \begin{center}
      \includegraphics[width=80mm]{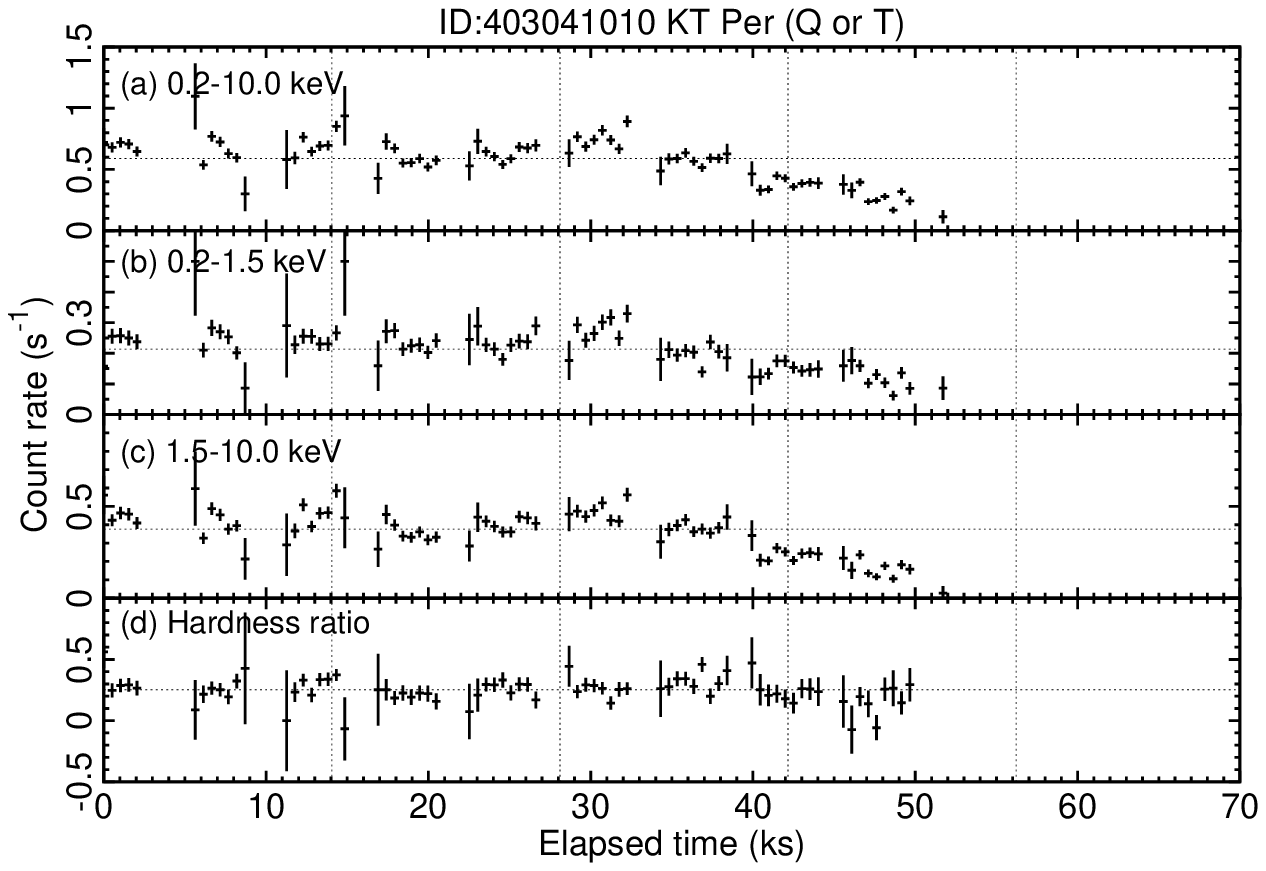}
    \end{center}
  \end{minipage}
  \begin{minipage}{0.5\hsize}
    \begin{center}
      \includegraphics[width=80mm]{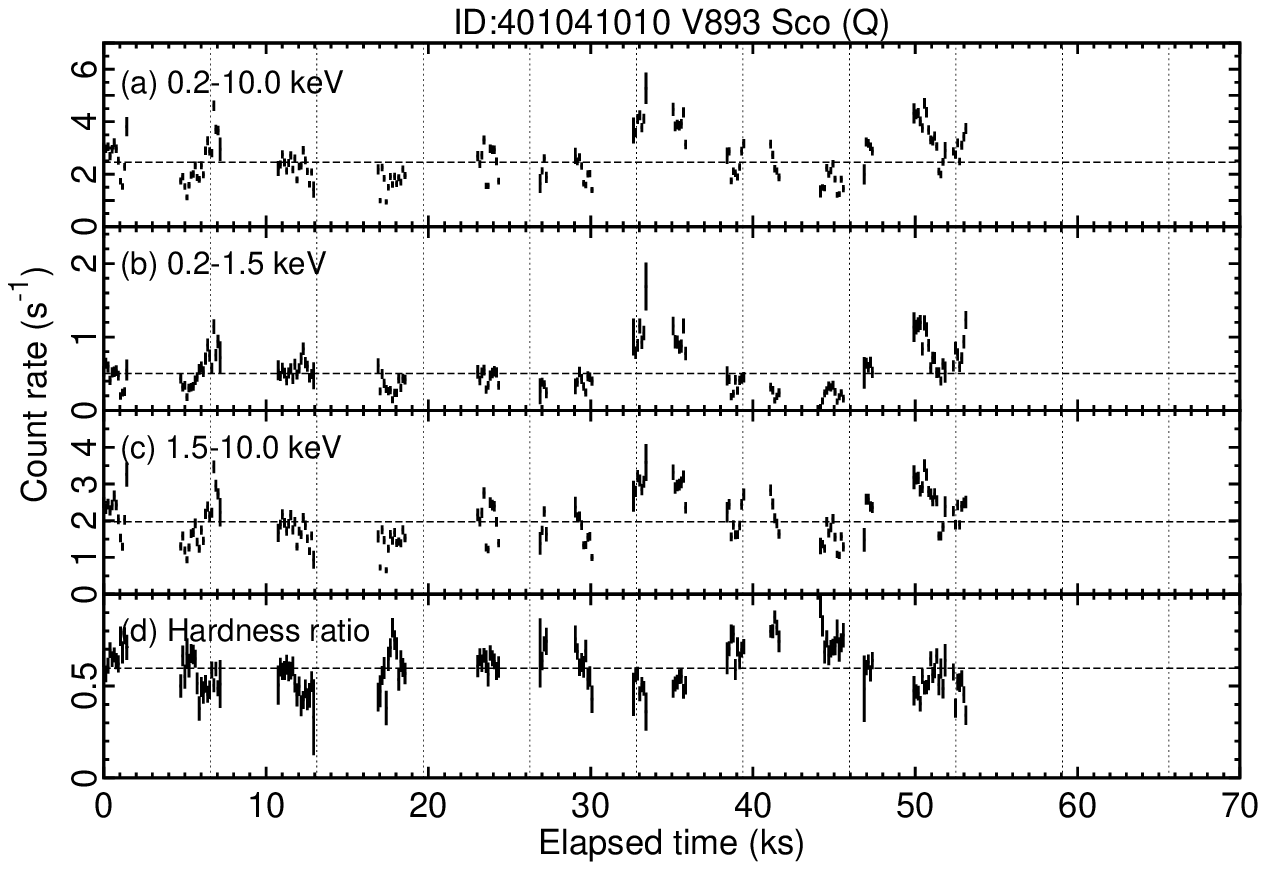}
    \end{center}
  \end{minipage}
  \vspace{1mm}
  \caption{Count rate and hardness ratio light curves. (a)~0.2--10.0~keV band,
    (b)~0.2--1.5~keV band (soft band; $S$), (c)~1.5--10.0~keV (hard band; $H$), and
    (d)~hardness ratio defined as $(H-S)/(H+S)$. In each panel, the horizontal and
    vertical lines indicate the median count rate and the orbital period,
    respectively.
  }\label{f8}
\end{figure*}

Possible explanations for the partial covering material include (i) a part of the
accretion disk interrupts the X-ray emission from the BL, or (ii) a part of the X-ray
emission is absorbed by an intervening matter such as clumpy disk wind. If the partial
covering absorber is a part of the disk, the variation is expected to be related to the
orbital period. We constructed light curves in the three energy bands (0.2--10.0,
0.2--1.5, and 1.5--10.0~keV band) and the hardness ratios in figure~\ref{f8} for these
sources. We do not see a clear variation synchronized with the orbital period, which
suggests that the explanation (ii) is more likely. This is consistent with
\citet{saitou12} using the same data set for Z Cam in the transitional state.

It is interesting to note that the Z Cam spectrum in the outburst state, which is
newly presented in this work, does not require an additional partial extinction. This
is in contrast with the ASCA result \citep{baskill05}, in which a partial covering
was required during the outburst state. The disk wind in Z Cam may be prominent only
during a part of the transitional and outburst states.

It is also noteworthy that an additional partial extinction was required also in two
sources in the quiescent state. For one of them (V893~Sco), \citet{mukai09} argued that
the origin of the intrinsic absorber is in the inner part of the disk, which is likely
for a high-inclination systems. The other source, BV~Cen, may be explained by the same
interpretation, though its inclination is not very high (table~\ref{t1}).

\citet{mukai09} also found a dip in the X-ray light curve of V893~Sco, which does not
accompany the change of the spectral shape. They attribute the dip to the partial
covering by the companion star. In our data, KT Per have a similar behavior; it exhibits
little change in the hardness ratio light curve unlike the other these sources, yet
shows a clear decline in the last quarter of the X-ray light curve (figure~\ref{f8}). The
decline does not repeat by the orbital phase, so the partial covering should be by
something else. There may be several different origins, and a multiple of them are
combined, for the additional extinction in DNe.

During the outbursts and super-outburst states, additional low-temperature thermal
plasma emission was required to explain the soft excess below 2~keV (\S~\ref{s3-4}
and \ref{s3-5}), which we consider to originate from a region different from the BL,
such as a corona around the WD and/or the disk.

During the super-outburst of VW~Hyi, a significant change of $T_{\rm min}$ was
observed, which may indicate that the plasma structure changed in the BL during this
state. This can be explained if the inner edge of the accretion disk becomes so dense
that the WD surface is beyond the $\tau=1$ depth seen from the observer. As a
consequence, $T_{\rm min}$, which is the BL temperature at $\tau\sim1$, becomes higher
than the WD surface temperature.

\section{Conclusions}\label{s5}
We carried out a systematic analysis of 21 data sets of the 15 DNe observed by
Suzaku. We classified DNe based on the optical light curves into the four states;
quiescent, transitional, outburst, and super-outburst states. We revealed a general
picture for each state in the X-rays.

For spectral analysis, we defined the isobaric cooling flow model with a Gaussian model
at 6.4~keV attenuated by a photoelectric absorption as the ``fiducial model''.  All the
X-ray spectra except two in the quiescent state were represented by the fiducial
model. This result suggests, as \citet{baskill05} and \citet{pandel05} pointed out, the
X-ray emission is radiated from the optically-thin multi-temperature plasma located in
the boundary layer in this state. For one DN in the transition state and another
possibly in the transition state, we fitted their spectra successfully with the fiducial
model modified with a partial covering absorption by neutral matter, which we consider
to be an intervening matter such as clumpy disk wind. However, the additional covering
can be attributed by other mechanisms such as a partial eclipse by the inner part of
the disk, the companion star, or something else. The outburst and super-outburst spectra
can be fitted with the fiducial model only in the 2.0--10.0~keV band. We found a
significant soft excess below 2.0~keV, which was reproduced by one or two thermal plasma
components. We speculate that a possible origin for the soft excess emission is a corona
somewhere in the system.  Furthermore, in the super-outburst state, it was necessary to
raise the minimum temperature of the white-dwarf surface in the fiducial model,
suggesting that the plasma structure changes in this state.

\bigskip

We appreciate the critique by the anonymous referee for improving the manuscript.
The authors are financially supported by the MEXT/JSPS KAKENHI Grant Numbers
JP14J11810 (Q.\,W.), JP24105007, JP15H03642, and JP16K05309 (M.\,T.), JP16K05309
(K.\,E.), and JP15J10520 and JP26800113 (T.\,H.). We acknowledge the variable star
observations from the AAVSO International Database contributed by observers
worldwide. This research made use of data obtained from Data ARchives and
Transmission System (DARTS), provided by Center for Science-satellite Operation and
Data Archives (C-SODA) at ISAS/JAXA, and the software tools provided by HEASARC at
NASA/GSFC.

\DeclareAbbreviation\na{NewA}
\bibliographystyle{aa}
\bibliography{ms}

\end{document}